\documentstyle[aps,eqsecnum,preprint,floats,epsf,epsfig]{revtex}
\begin{document}
\def\be{\begin{eqnarray}}
\def\en{\end{eqnarray}}
\def\non{\nonumber}
\def\la{\langle}
\def\ra{\rangle}
\def\nc{N_c^{\rm eff}}
\def\vp{\varepsilon}
\def\A{{\cal A}}
\def\B{{\cal B}}
\def\up{\uparrow}
\def\dw{\downarrow}
\def\vma{{_{V-A}}}
\def\vpa{{_{V+A}}}
\def\smp{{_{S-P}}}
\def\spp{{_{S+P}}}
\def\J{{J/\psi}}
\def\ov{\overline}
\def\Lqcd{{\Lambda_{\rm QCD}}}
\def\pr{{\sl Phys. Rev.}~}
\def\prl{{\sl Phys. Rev. Lett.}~}
\def\pl{{\sl Phys. Lett.}~}
\def\np{{\sl Nucl. Phys.}~}
\def\zp{{\sl Z. Phys.}~}
\def\lsim{ {\ \lower-1.2pt\vbox{\hbox{\rlap{$<$}\lower5pt\vbox{\hbox{$\sim$}
}}}\ } }
\def\gsim{ {\ \lower-1.2pt\vbox{\hbox{\rlap{$>$}\lower5pt\vbox{\hbox{$\sim$}
}}}\ } }

\font\el=cmbx10 scaled \magstep2{\obeylines\hfill May, 2002}

\vskip 1.5 cm

\centerline{\large\bf Charmless Exclusive Baryonic $B$ Decays}
\bigskip
\centerline{\bf Hai-Yang Cheng$^{1,2}$ and Kwei-Chou Yang$^{3}$}
\medskip
\centerline{$^1$ Institute of Physics, Academia Sinica}
\centerline{Taipei, Taiwan 115, Republic of China}
\medskip
\centerline{$^2$ C.N. Yang Institute for Theoretical Physics,
State University of New York} \centerline{Stony Brook, New York
11794}
\medskip
\centerline{$^3$ Department of Physics, Chung Yuan Christian
University} \centerline{Chung-Li, Taiwan 320, Republic of China}
\bigskip
\bigskip
\centerline{\bf Abstract}
\bigskip
{\small We present a systematical study of two-body and three-body
charmless baryonic $B$ decays. Branching ratios for two-body modes
are in general very small, typically less than $10^{-6}$, except
that $\B(B^-\to p \bar\Delta^{--})\sim 1\times 10^{-6}$. In
general, $\ov B\to N\bar\Delta>\ov B\to N\bar N$ due to the large
coupling constant for $\Sigma_b\to B\Delta$. For three-body modes
we focus on octet baryon final states. The leading three-dominated
modes are $\ov B^0\to p\bar n\pi^-(\rho^-),~n\bar p\pi^+(\rho^+)$
with a branching ratio of order $3\times 10^{-6}$ for $\ov B^0\to
p\bar n\pi^-$ and $8\times 10^{-6}$ for $\ov B^0\to p\bar
n\rho^-$.  The penguin-dominated decays with strangeness in the
meson, e.g., $B^-\to p\bar p K^{-(*)}$ and $\ov B^0\to p\bar n
K^{-(*)},~n\bar n \bar K^{0(*)}$, have appreciable rates and the
$N\bar N$ mass spectrum peaks at low mass. The penguin-dominated
modes containing a strange baryon, e.g., $\ov B^0\to \Sigma^0\bar
p\pi^+,~\Sigma^-\bar n\pi^+$, have branching ratios of order
$(1\sim 4)\times 10^{-6}$. In contrast, the decay rate of $\ov
B^0\to\Lambda\bar p\pi^+$ is smaller. We explain why some of
charmless three-body final states in which baryon-antibaryon pair
production is accompanied by a meson have a larger rate than their
two-body counterparts: either the pole diagrams for the former
have an anti-triplet bottom baryon intermediate state, which has a
large coupling to the $B$ meson and the nucleon, or they are
dominated by the factorizable external $W$-emission process.

}

\pagebreak

\section{Introduction}
Inspired by the claim of the observation of the decay modes $p\bar
p\pi^\pm$ and $p\bar p\pi^+\pi^-$ in $B$ decays by ARGUS
\cite{ARGUS} in the late 1980s, baryonic $B$ decays were studied
extensively around the early 1990s
\cite{DTS,Paver,Gronau,Korner,Dobr,Chernyak,He,Lu,Jarfi,Ball,Khanna,Kaur}
with the focus on the tree-dominated two-body decay modes, e.g.
the charmful decays $B\to\Lambda_c\bar N,~\Sigma_c\bar N$, and
charmless ones $B\to p\bar p,~\Lambda\bar\Lambda$. Up to now, none
of the two-body baryonic $B$ decays have been observed
\cite{CLEOa,Belle}. Many of the earlier model predictions are too
large compared to experiment. For example, the previous limit on
$\ov B^0\to p\bar p<7\times 10^{-6}$ set by CLEO \cite{CLEOa} has
been recently pushed down to the level of $1.6\times 10^{-6}$ by
Belle \cite{Belle}, whereas the model predictions are either too
large or marginally comparable to the experimental limit (see
Table II below).

The penguin-induced charmless baryonic $B$ decays such as $\ov
B\to \Sigma\bar p,~\Sigma\bar \Delta$ have been studied by
Chernyak and Zhitnitsky \cite{Chernyak} based on the QCD sum rule
analysis. They obtained the branching ratios of order
$(0.3-1.0)\times 10^{-5}$. Experimentally, only the upper limits
on $B^-\to \Lambda\bar p,~\Lambda\bar p\pi^+\pi^-,~\Delta^0\bar
p,~ p\bar\Delta^{--}$ ($\bar\Delta^{--}$ being the antiparticle of
$\Delta^{++}$) and $\ov B^0\to\Lambda\bar p\pi^+$ have been set.

As pointed out by Dunietz \cite{Dunietz} and by Hou and Soni
\cite{HS}, the smallness of the two-body baryonic decay
$B\to\B_1\ov\B_2$ has to do with a straightforward Dalitz plot
analysis (see Sec. IV for a detailed discussion) or with the large
energy release. Hou and Soni conjectured that in order to have
larger baryonic $B$ decays, one has to reduce the energy release
and  at the same time allow for baryonic ingredients to be present
in the final state. Under this argument, the three-body decay, for
example $B\to \rho p\bar n$, will dominate over the two-body mode
$B\to p\bar n$ since the ejected $\rho$ meson in the former decay
carries away much energies and the configuration is more favorable
for baryon production because of reduced energy release compared
to the latter \cite{CHT2}. This is in contrast to the mesonic $B$
decays where the two-body decay rate is generally comparable to
the three-body one. The large rate of $B^0\to D^{*-}p\bar n$ and
$B^0\to D^{*-}p\bar p\pi^+$ observed by CLEO \cite{CLEOb}
indicates that the decays $B\to$ baryons receive comparable
contributions from $\ov B\to\Lambda_c \bar p X$ and $\ov B\to
DN\bar N' X$, as originally advocated by Dunietz \cite{Dunietz}. A
theoretical study of the decay $B\to D^* p\bar n$ has been carried
out recently by \cite{CHT1}. In \cite{CYBbaryon} we have shown
explicitly that the three-body charmful decay $B^-\to\Lambda_c\bar
p\pi^-(\rho^-)$ has indeed a magnitude larger than $\ov
B^0\to\Lambda_c\bar p$ as seen experimentally \cite{CLEOc}. By the
same token, it is expected that for charmless baryonic $B$ decays,
$\ov B\to (\pi,\rho)\B_1\ov\B_2$ are the dominant modes induced by
tree operators and $\ov B\to(\pi,\rho)\B_{1(s)}\ov \B_2$, $\ov
B\to K^{(*)}\B_1\ov \B_2$ are the leading modes induced by penguin
diagrams. The recent first observation of the penguin-dominated
charmless baryonic decay $B^-\to p\bar p K^-$ by Belle
\cite{Bellebaryon} clearly indicates that it has a much larger
rate than the two-body counterpart $\ov B^0\to p\bar p$. Of
course, this does not necessarily imply that the three-body final
state $\B_1\ov \B_2M$ always has a branching ratio larger than the
two-body one $\B_1\ov\B_2$. We shall examine under what
circumstance that the above argument holds.

In the present paper we will give a systematical study of two-body
and three-body charmless baryonic $B$ decays. The factorizable
$W$-exchange or $W$-annihilation contribution to two-body decay
modes is very small and hence negligible. For nonfactorizable
contributions to two-body final states, we will calculate the
corresponding pole diagrams at the hadron level. We will apply the
bag model to evaluate the baryon-baryon matrix elements and find
that the baryon-strange baryon weak transition is indeed dominated
by penguin operators. Branching ratios for two-body baryonic modes
are found to be in general very small $\lsim {\cal O}(10^{-6})$
except for the decays with a $\Delta$ resonance in the final
state.

The study of three-body baryonic decays is more complicated.
Though it in general receives factorizable contributions, some of
them involve three-body matrix elements and hence are not ready to
evaluate. Therefore, pole diagrams still play an essential role.
The baryonic decay with a vector meson in the final state normally
has a large rate which should be easily accessible by the existing
$B$ factories.

The layout of the present paper is organized as follows. In Sec.
II the issue of renormalization scheme and scale dependence of
Wilson coefficients is addressed. We then study charmless two-body
baryonic $B$ decays in Sec. III and compare our results with the
literature and experiment. In Sec. IV some important three-body
modes are analyzed.  Sec. V gives discussions and conclusions. A
short summary of the relevant baryon wave functions and the bag
model evaluation of baryon-baryon matrix elements are presented in
the Appendix.

\section{Hamiltonian}
The relevant effective $\Delta B=1$ weak Hamiltonian for hadronic
charmless $B$ decays is \be {\cal H}_{\rm eff}(\Delta B=1) &=&
{G_F\over\sqrt{2}}\Bigg\{ V_{ub}V_{uq}^*
\Big[c_1(\mu)O_1^u(\mu)+c_2(\mu)O_2^u(\mu)\Big]+V_{cb}V_{cq}^*\Big[c_1(\mu)
O_1^c(\mu)+c_2(\mu)O_2^c(\mu)\Big]  \non \\
&& -V_{tb}V_{tq}^*\sum^{10}_{i=3}c_i(\mu)O_i(\mu)\Bigg\}+{\rm
h.c.}, \en where $q=d,s$, and \be && O_1^u= (\bar ub)_\vma(\bar
qu)_\vma, \qquad\qquad\qquad\qquad~~
O_2^u = (\bar u_\alpha b_\beta)_\vma(\bar q_\beta u_\alpha)_\vma, \non \\
&& O_1^c= (\bar cb)_\vma(\bar qc)_\vma,
\qquad\qquad\qquad\qquad~~~
O_2^c = (\bar c_\alpha b_\beta)_\vma(\bar q_\beta c_\alpha)_\vma, \non \\
&& O_{3(5)}=(\bar qb)_\vma\sum_{q'}(\bar q'q')_{\vma(\vpa)},
\qquad  \qquad O_{4(6)}=(\bar q_\alpha b_\beta)_\vma\sum_{q'}(\bar
q'_\beta q'_\alpha)_{
\vma(\vpa)},   \\
&& O_{7(9)}={3\over 2}(\bar qb)_\vma\sum_{q'}e_{q'}(\bar
q'q')_{\vpa(\vma)},
  \qquad O_{8(10)}={3\over 2}(\bar q_\alpha b_\beta)_\vma\sum_{q'}e_{q'}(\bar
q'_\beta q'_\alpha)_{\vpa(\vma)},   \non \en with $O_3$--$O_6$
being the QCD penguin operators, $O_{7}$--$O_{10}$ the electroweak
penguin operators and $(\bar q_1q_2)_{_{V\pm A}}\equiv\bar
q_1\gamma_\mu(1\pm \gamma_5)q_2$. The scale dependent Wilson
coefficients calculated at next-to-leading order are
renormalization scheme dependent. We use the next-to-leading
Wilson coefficients evaluated in the naive dimensional
regularization scheme \cite{Buras96}
 \be
c_1=1.082, \quad c_2=-0.185,\quad c_3=0.014, \quad c_4=-0.035,
\quad c_5=0.009, \quad c_6=-0.041, \non \\
c_7/\alpha=-0.002, \quad c_8/\alpha=0.054, \quad
c_9/\alpha=-1.292, \quad c_{10}/\alpha=0.263,\quad c_g=-0.143,
 \en
at $\mu=\overline{ m}_b(m_b)=4.40$ GeV for
$\Lambda^{(5)}_{\overline{\rm MS}}=225$ MeV taken from Table XXII
of \cite{Buras96} with $\alpha$ being an electromagnetic
fine-structure coupling constant. In order to ensure that the
physical amplitude is renormalization scale and $\gamma_5$-scheme
independent, we include vertex and penguin corrections to hadronic
matrix elements of four-quark operators \cite{CCTY,CY00}. This
amounts to modifying $c_i(\mu) \to c_i^{\rm eff}$ and
 \be
 \sum  c_i(\mu)\la Q_i(\mu)\ra=\sum c_i^{\rm eff}\la Q_i\ra_{\rm
 VIA},
 \en
where the subscript VIA means that the hadronic matrix element is
evaluated under the vacuum insertion approximation.  Numerical
results for $c_i^{\rm eff}$ are shown in Table I (for details, see
\cite{CCTY}). It should be stressed that $c_i^{\rm eff}$ are
renormalization scale and scheme independent. For the mesonic
decay $B\to M_1M_2$ with two mesons in the final state, two of the
four quarks involving in the vertex diagrams will form an ejected
meson. In this case, it is necessary to take into account the
convolution with the ejected meson wave function.

The penguin matrix element of scalar and pseudoscalar densities,
$\la \B_1\ov \B_2|\bar q_1(1\pm\gamma_5)q_2|0\ra$, is usually
evaluated by applying the equation of motion and it is
renormalization scale and scheme dependent. Since the
factorization scale is set at $\mu_f=m_b$ to obtain the effective
Wilson coefficients listed in Table I, we will therefore evaluate
the penguin matrix element of scalar and pseudoscalar densities at
the $m_b$ scale.

\vskip 0.4cm
\begin{table}[ht]
\caption{Numerical values of the effective Wilson coefficients
$c_i^{\rm eff}$ for $b\to s$, $b\to d$ and $\bar b\to\bar d$
transitions evaluated at $\mu_f=m_b$ and $k^2=m_b^2/2$ taken from
Table I of [26], where use of $|V_{ub}/V_{cb}|=0.085$ has been
made. The numerical results are insensitive to the unitarity angle
$\gamma$.}
\begin{center}
\begin{tabular}{ l c c c  }
 & $b\to s$,~$\bar b\to\bar s$ & $b\to d$ & $\bar b\to\bar d$  \\ \hline
$c_1^{\rm eff}$ & 1.169 & 1.169 & 1.169 \\ $c_2^{\rm eff}$ &
$-0.367$ & $-0.367$ & $-0.367$ \\ $c_3^{\rm eff}$ &
$0.0227+i0.0045$ & $0.0226+i0.0038$ & $0.0230+i0.0051$ \\
$c_4^{\rm eff}$ & $-0.0463-i0.0136$ & $-0.0460-i0.0114$ &
$-0.0470-i0.0154$
\\ $c_5^{\rm eff}$ & $0.0134+i0.0045$ & $0.0133+i0.0038$ &
$0.0137+i0.0051$ \\
 $c_6^{\rm eff}$ & $-0.0600-i0.0136$ &
$-0.0597-i0.0114$ & $-0.0608-i0.0154$ \\
 $c_7^{\rm eff}/\alpha$ &
$-0.0309-i0.0367$ & $-0.0305-i0.0324$ & $-0.0326-i0.0403$ \\
$c_8^{\rm eff}/\alpha$ & 0.070 & 0.070 & 0.070 \\ $c_9^{\rm
eff}/\alpha$ & $-1.428-i0.0367$ & $-1.428-i0.0324$ &
$-1.430-i0.0403$
\\ $c_{10}^{\rm eff}/\alpha$ & 0.48 & 0.48 & 0.48
\\
\end{tabular}
\end{center}
\end{table}

For quark mixing matrix elements, we will use
$|V_{ub}/V_{cb}|=0.085$ and the unitary angle $\gamma=60^\circ$.
In terms of the Wolfenstein parameters $A=0.815$ and
$\lambda=0.2205$ we have
 \be
 \rho=0.385\,\sin\gamma, \qquad\quad \eta=0.385\,\cos\gamma,
 \en
where $\rho$ and $\eta$ are the parameters in the Wolfenstein
parametrization  \cite{Wolf} of the quark mixing matrix.

\section{charmless two-body baryonic decays}
The charmless $B$ decays into two light baryons can be classified
into two categories: the ones induced by the $b\to u$ tree
transition, and the ones by the $b\to s$ penguin transition. The
decay amplitude of $B\to \B_1({1\over 2}^+)\ov \B_2({1\over 2}^+)$
has the form
 \be
 {\cal A}(B\to \B_1\ov \B_2)=\bar u_1(A+B\gamma_5)v_2,
 \en
where $A$ and $B$ correspond to $p$-wave parity-violating (PV) and
$s$-wave parity-conserving (PC) amplitudes, respectively. The
decay rate is given by
 \be
 \Gamma(B\to \B_1({1/ 2}^+)\ov \B_2({1/ 2}^+))&=& {p_c\over 4\pi}\Bigg\{
 |A|^2\,{(m_B+m_1+m_2)^2p_c^2\over (E_1+m_1)(E_2+m_2)m_B^2}\non \\
 & +&
 |B|^2\,{[(E_1+m_1)(E_2+m_2)+p_c^2]^2\over
 (E_1+m_1)(E_2+m_2)m_B^2} \Bigg\},
 \en
where $p_c$ is the c.m. momentum, $E_i$ and $m_i$ are the energy
and mass of the baryon $\B_i$, respectively. For the decay $B\to
\B_1({3\over 2}^+)\ov \B_2({1\over 2}^+)$ with a spin-${3\over 2}$
baryon in the final state, the general amplitude reads
 \be
 \A(B\to \B_1(p_1)\ov \B_2(p_2))=iq_\mu\bar u^\mu_1(p_1)(C+D\gamma_5)v_2(p_2),
 \en
where $u^\mu$ is the Rarita-Schwinger vector spinor for a
spin-${3\over 2}$ particle, $q=p_1-p_2$ and $C,~D$ correspond to
parity-violating $p$-wave and parity-conserving $d$-wave
amplitudes, respectively. The corresponding decay rate is
 \be
 \Gamma(B\to \B_1({3/ 2}^+)\ov \B_2({1/ 2}^+))&=& {p_c^3\over 6\pi}\,{1\over m_1^2}\Bigg\{
 |C|^2\,{[(E_1+m_1)(E_2+m_2)+p_c^2]^2\over
 (E_1+m_1)(E_2+m_2)m_B^2} \non \\
 & +& |D|^2\,{(m_B+m_1+m_2)^2p_c^2\over (E_1+m_1)(E_2+m_2)m_B^2}
  \Bigg\}.
 \en

As shown in Fig. 1, the quark diagrams for two-body baryonic $B$
decays consist of  internal $W$-emission diagram, $b\to d(s)$
penguin transition, $W$-exchange for the neutral $B$ meson and
$W$-annihilation for the charged $B$. Just as mesonic $B$ decays,
$W$-exchange and $W$-annihilation are expected to be helicity
suppressed and the former is furthermore subject to color
suppression.\footnote{In contrast, $W$-exchange plays an essential
role in nonleptonic decays of baryons as it is no longer subject
to color and helicity suppression.} In the language of the pole
model, the $M^{(*)}\B_1\ov\B_2$ form factor is expected to be
largely suppressed at $q^2=m_B^2$. As estimated by
\cite{Korner,Jarfi,Kaur}, the $W$-exchange or $W$-annihilation
contribution is very insignificant and hence can be neglected. The
tree-dominated decays, e.g. $\ov B^0\to p\bar p,~p\bar\Delta^-$
are mainly induced by the internal $W$-emission via $b\to u$
transition, while penguin-dominated modes, e.g. $B^-\to\Lambda
\bar p,~\Sigma^0\bar p$ proceed through $b\to s$ penguin
transition. These amplitudes are nonfactorizable and thus very
difficult to evaluate directly. This is the case in particular for
baryons, which being made out of three quarks, in contrast to two
quarks for mesons, bring along several essential complications. In
order to circumvent this difficulty, it is customary to assume
that the decay amplitude at the hadron level is dominated by the
pole diagrams with low-lying one-particle intermediate states.
More precisely, PC and PV amplitudes are dominated by ${1\over
2}^+$ ground-state intermediate states and ${1\over 2}^-$
low-lying baryon resonances, respectively
\cite{Jarfi}.\footnote{The $s$-channel meson pole states
correspond to weak annihilation diagrams [see Fig. 1(b)].} This
pole model has been applied successfully to nonleptonic decays of
hyperons and charmed baryons \cite{CT92,CT93}. In general, the
pole diagram leads to
 \be
 A=-\sum_{\B_b^*}{g_{\B_b^{*}\to B\B_2}\,b_{\B_b^*\B_1}\over
 m_{1}-m_{\B_b^*} }, \qquad\qquad B=\sum_{\B_b}{g_{\B_b\to
 B\B_2}\,
 a_{\B_b\B_1}\over m_{1}-m_{\B_b}},
 \en
where
 \be
 \la \B_1|{\cal H}_{\rm eff}^{\rm PC}|\B_b\ra = \bar
 u_{\B_1}a_{\B_b\B_1}u_{\B_b}, \qquad\quad
 \la \B_1|{\cal H}_{\rm eff}^{\rm PV}|\B_b^{*}\ra = i\bar
 u_{\B_1}b_{\B_b^*\B_1}u_{\B_b^*}
 \en
are PC and PV matrix elements, respectively.

Since the weak transition does not involve momentum transfer, it
can be evaluated using the quark model. Conventionally, baryon
matrix elements are evaluated using the bag model or the harmonic
oscillator model. In the present work we prefer to employ the MIT
bag model \cite{MIT} to compute the weak baryon-baryon transition
for several reasons. First, it has been applied successfully to
describe the $p$-wave amplitudes of hyperon nonleptonic decays and
it is much simpler than the harmonic oscillator model for
computing the PC matrix elements. Second, it is relatively easy to
incorporate penguin operators in calculations. Third and most
importantly, the bag model calculation gives predictions
consistent with experiment, whereas the calculated results based
on the harmonic-oscillator model are too large compared to the
data. This will be clearly demonstrated below when we discuss
$B\to p\bar p$ and $p\bar\Delta$.

However, it is known that the bag model is considerably less
successful for describing the physical non-charm and non-bottom
${1\over 2}^-$ resonances \cite{MIT}, not mentioning the charm or
bottom ${1\over 2}^-$ baryon states. Therefore, we will not
evaluate the PV matrix element $b_{\B^*\B}$ and the strong
coupling $g_{\B_b^{*}\to B\B_2}$ as their calculations in the bag
model are much more involved and are far more uncertain than the
PC case \cite{CT92}. Fortunately, there are some decay modes that
are purely parity-conserving within the framework of the $^3P_0$
quark-pair-creation model to be mentioned shortly. Examples are
$B^-\to n\bar p$ and $\ov B\to N\bar\Delta$, which will be
discussed below.

\begin{figure}[ptb]
\hspace{4cm} \psfig{figure=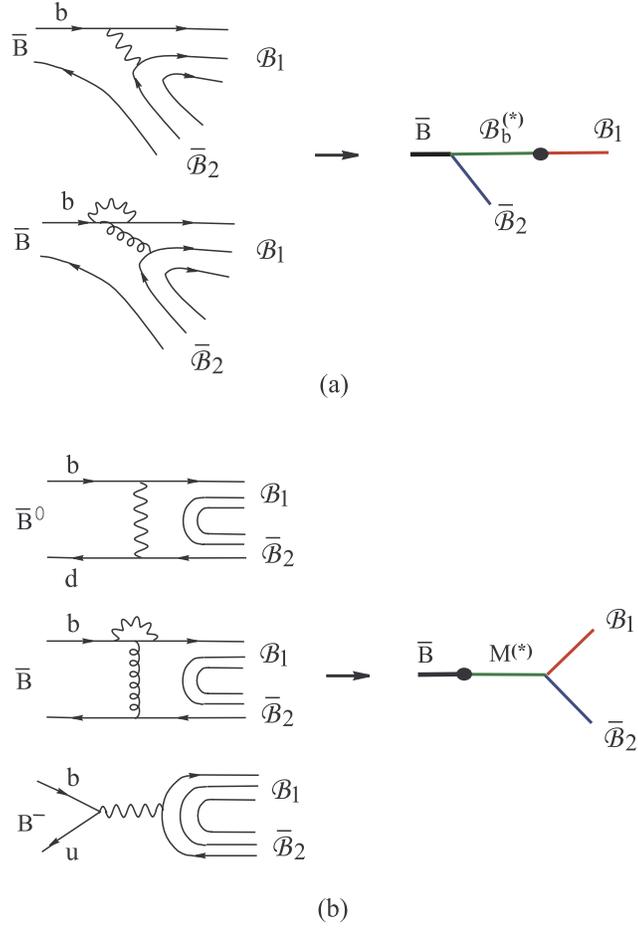,height=5in} \vspace{1.2cm}
    \caption{{\small Quark and pole diagrams for $\ov B\to\B_1\ov \B_2$
    where the symbol $\bullet$ denotes the weak vertex. Fig. 1(a) corresponds
    to the nonfactorizable internal $W$ emission or the $b\to d(s)$ penguin transition,
    while Fig. 1(b) to the $W$-exchange contribution
    for the neutral $B$ or $W$-annihilation for the charged $B$, or
    penguin-induced weak annihilation.
    }}
   \label{fig:1}
\end{figure}

For strong couplings we will follow \cite{Jarfi,Yaouanc} to adopt
the $^3P_0$ quark-pair-creation model in which the $q\bar q$ pair
is created from the vacuum with vacuum quantum numbers $^3P_0$. We
shall apply this model to estimate the relative strong coupling
strength and choose $|g_{\Sigma_b^+\to\ov B^0p}|=5$ as a
benchmarked value for the absolute coupling strength (see below).
Presumably, the $^3P_0$ model works in the nonperturbative low
energy regime. In contrast, in the perturbative high energy region
where perturbative QCD is applicable, it is expected that the
quark pair is created perturbatively via one gluon exchange with
one-gluon quantum numbers $^3S_1$. Since the light baryons
produced in two-body baryonic $B$ decays are very energetic, it
appears that the $^3S_1$ model may be more relevant. However, in
the present paper we adopt the $^3P_0$ model for quark pair
creation for the following two reasons. First, it is much simpler
to estimate the relative strong coupling strength in the $^3P_0$
model  rather than in the $^3S_1$ model where hard gluons arise
from four different quark legs and generally involve infrared
problems. Second, this model is presumably reliable for estimating
the $\B_b\B B$ coupling when all particles are on their mass
shell. Of course, the intermediate pole state $\B_b$ in the
two-body baryonic decay is far from its mass shell (but not quite
so in the three-body decay). In principle, one can treat the
intermediate state as an on-shell particle and then assume that
off-shell effects of the pole can be parametrized in terms of form
factors. Such form factors are basically unknown, though they are
expected to become smaller as the intermediate state is more away
from its mass shell due to less overlap of initial and final
hadron wave functions. Since we are interested in the relative
strength of strong couplings rather than the absolute strength, it
seems plausible to assume that the relative coupling strengths are
essentially not affected by the off-shell extrapolation; that is,
the strong form factors are assumed to be universal. We then use
the experimental result for $B^-\to\Lambda_c\bar p\pi^-$ to fix
the absolute coupling strength of $g_{\Lambda_b^+\to\ov B^0p}$ or
$g_{\Sigma_b^+\to\ov B^0p}$ \cite{CYBbaryon}.\footnote{The
nonresonant decay $B^-\to\Lambda_c\bar p\pi^-$ receives main
contributions from Figs. 2(a) and 2(d) shown in Sec. IV (see
\cite{CYBbaryon}). In the pole model, the contribution of the
former is governed by the $\Lambda_b$ pole. Therefore, a
measurement of the decay rate of this mode enables us to determine
the off-shell coupling $g_{\Lambda_b^+\to\ov B^0p}$. For detail,
see \cite{CYBbaryon}.} In the future, it is important to carry out
the more sophisticated pQCD analysis to gain a better
understanding of the underlying decay mechanism for baryonic $B$
decays.

At this point, we would like to stress that although we employ the
same pole-model framework as Jarfi {\it et al.} \cite{Jarfi} to
discuss baryonic $B$ decays, the calculational detail is
different. While Jarfi {\it et al.} evaluated baryon matrix
elements at large momentum transfer and strong couplings at small
transfer, we consider weak transition at zero transfer and strong
couplings at large momentum transfer as elaborated before. Another
difference is related to the quark model evaluation of baryon
matrix elements: We employ the bag model rather than the harmonic
oscillator model.

For the reader's convenience, in Table II we give a summary of the
calculational results presented in Sections III.A and III.B below.
For comparison, some other predictions in the literature are shown
in the same table.

\begin{table}[t]
\caption{Predictions of the branching ratios for some charmless
two-body baryonic $B$ decays classified into two categories:
tree-dominated and penguin-dominated. In this work, some branching
ratios denoted by ``$\dagger$" are calculated only for the
parity-conserving part. For comparison some other predictions in
the literature are also shown. We have normalized the branching
ratios to $|V_{ub}/V_{cb}|=0.085\,$. The predictions given in [11]
are carried out in two different quark-pair-creation models: local
and nonlocal. The line separates tree- and penguin-dominated
charmless baryonic $B$ decays and experimental limits are taken
from [14,15].}
\begin{center}
\begin{tabular}{l| c c c c c c c }
&  &  &  &  \multicolumn{2}{c}{Ref.[11]} & &    \\ \cline{5-6}
\raisebox{1.5ex}[0cm][0cm]{} &
\raisebox{1.5ex}[0cm][0cm]{Ref.\cite{Paver}} &
\raisebox{1.5ex}[0cm][0cm]{Ref.\cite{Chernyak}} &
\raisebox{1.5ex}[0cm][0cm]{Ref.\cite{Jarfi}}
 & non-local & local &  \raisebox{1.5ex}[0cm][0cm]{This work} &
 \raisebox{1.5ex}[0cm][0cm]{Expt.}  \\ \hline
 $\ov B^0\to p\bar p$ & $4.2\times 10^{-6}$ & $1.2\times 10^{-6}$ & $7.0\times
 10^{-6}$  & $2.9\times 10^{-6}$ & $2.7\times 10^{-5}$ & $1.1\times 10^{-7\dagger}$
 & $<1.2\times  10^{-6}$ \\
 $\ov B^0\to n\bar n$ & & $3.5\times 10^{-7}$ & $7.0\times
 10^{-6}$ & $2.9\times 10^{-6}$ & $2.7\times 10^{-5}$ & $1.2\times
 10^{-7\dagger}$ & \\
 $B^-\to n\bar p$ & & $6.9\times 10^{-7}$ & $1.7\times
 10^{-5}$ & 0 & 0 & $5.0\times  10^{-7}$ & \\
 $\ov B^0\to\Lambda\bar\Lambda$ & & & $2\times 10^{-7}$ & & & $0^\dagger$ &
 $<1.0\times 10^{-6}$ \\
 $B^-\to p\bar \Delta^{--}$ & $1.5\times 10^{-4}$ & $2.9\times 10^{-7}$
 &  $3.2\times 10^{-4}$ & $2.4\times 10^{-6}$ & $ 8.7\times 10^{-6}$
 &  $1.4\times 10^{-6}$ & $<1.5\times 10^{-4}$ \\
 $\ov B^0\to p\bar\Delta^-$ & & $7\times 10^{-8}$ & $1.0\times
 10^{-4}$ & $1.0\times 10^{-6}$ & $4.0\times 10^{-6}$ & $4.3\times
 10^{-7}$ & \\
 $B^-\to n\bar\Delta^-$ & &  & $1.1\times 10^{-4}$ & $2.7\times 10^{-7}$
 & $1\times 10^{-7}$ & $4.6\times  10^{-7}$ & \\
 $\ov B^0\to n\bar\Delta^0 $ & & & $1.0\times 10^{-4}$ & $1.0\times 10^{-6}$
 & $4.0\times 10^{-6}$ & $4.3\times 10^{-7}$ & \\
 \hline
 $B^-\to\Lambda\bar p$ & & $\lsim 3\times 10^{-6}$ & & & & $2.2\times 10^{-7\dagger}$ &
 $<2.2\times 10^{-6}$ \\
 $\ov B^0\to \Lambda\bar n$ & & & & & & $2.1\times 10^{-7\dagger}$ & \\
 $\ov B^0\to\Sigma^+\bar p$ & &  $6\times 10^{-6}$ & & & & $1.8\times
 10^{-8\dagger}$ & \\
 $B^-\to\Sigma^0\bar p$ & & $3\times 10^{-6}$ & & & & $5.8\times 10^{-8\dagger}$ & \\
 $B^-\to\Sigma^+\bar\Delta^{--}$ & & $6\times 10^{-6}$ & & & & $2.0\times 10^{-7}$ & \\
 $\ov B^0\to\Sigma^+\bar\Delta^-$ & & $6\times 10^{-6}$ & & & & $6.3\times 10^{-8}$ & \\
 $B^-\to\Sigma^-\bar\Delta^0$ & & $2\times 10^{-6}$ & & & & $8.7\times 10^{-8}$ & \\
\end{tabular}
\end{center}
\end{table}

\subsection{Tree-dominated two-body decays}
 \vskip 0.4 cm
\centerline{\underline{1.~~$\ov B^0\to p\,\bar p$}}
 \vskip 0.4cm
As discussed before, we can neglect $W$-exchange contributions to
$\ov B^0\to p\bar p$ and simply focus on the internal $W$-emission
which is manifested as the pole diagram at the hadron level with
the low-lying intermediate states $\Sigma_b^{+(*)}$  [see
Fig.1(a)]. The PV and PC wave amplitudes read
 \be
 A=-{g_{\Sigma_b^{+*}\to\ov B^0p}\,b_{\Sigma_b^{*+}p}\over
 m_p-m_{\Sigma_b^*} }, \qquad\qquad B=\,{g_{\Sigma_b^+\to\ov
 B^0p}\,  a_{\Sigma_b^+ p}\over
 m_p-m_{\Sigma_b}}.
 \en
Neglecting penguin contributions to the matrix element due to the
smallness of penguin coefficients, we have
 \be \label{PCm.e.}
 a_{\Sigma_b^+ p}={G_F\over
 \sqrt{2}}V_{ub}V_{ud}^*\,(c_1^{\rm eff}-c_2^{\rm eff}) \la
 p|O_1^{\rm PC}|\Sigma^+_b\ra
 \en
for the PC matrix element, where $O_1=(\bar ub)_\vma(\bar
du)_\vma$ and use has been made of $\la p|O_2|\Sigma^+_b\ra=-\la
p|O_1|\Sigma^+_b\ra$. The latter relation holds because the
combination of the four-quark operators $O_1+O_2$ is symmetric in
color indices (more precisely, it is a color sextet) and hence it
does not contribute to the baryon-baryon matrix element since the
baryon-color wave function is totally antisymmetric. In contrast,
the operator $O_1-O_2$ is a color antitriplet and has isospin
$I={1\over 2}$ because the diquark $ud$ is isoscalar due to
anti-symmetrization. The latter feature will lead to some $\Delta
I={1\over 2}$ rule relations, for example (\ref{isospin}) below.

We shall employ the MIT bag model \cite{MIT} to evaluate the
baryon matrix elements (see e.g. \cite{CT92,CT93} for the method).
From the Appendix of \cite{CYBbaryon} or \cite{CT92} we obtain the
PC matrix element
 \be \label{pSigmab}
 \la p|O_1^{\rm PC}|\Sigma^+_b\ra=-6X(4\pi),
 \en
where
 \be \label{bagX}
 X = \int^R_0 r^2dr[u_u(r)u_b(r)+v_u(r)v_b(r)][u_d(r)u_u(r)+v_d(r)v_u(r)],
 \en
is a four-quark overlap bag integral and $u_q(r)$, $v_q(r)$ are
the large and small components of the quark wave functions in the
ground $(1S_{1/2})$ state (see, for example, \cite{CYBbaryon}). As
stressed in passing, we will not evaluate the PV matrix element
$b_{\Sigma_b^*p}$ as its calculation in the bag model is much more
involved and considerably less reliable than the PC one. (However,
see \cite{CT92,CT93} for the evaluation of PV matrix elements in
charmed baryon decays.)  Numerically, we obtain
 \be
 X=1.52\times 10^{-4}\,{\rm GeV}^3.
 \en
Collecting everything together leads to
 \be
 \B(\ov B^0\to p\bar p)_{\rm PC}=1.1\times
 10^{-7}\left|{g_{\Sigma_b^+\to\ov B^0p}\over 5}\right|^2,
 \en
and hence
 \be
 \B(\ov B^0\to p\bar p)\lsim 2.2\times
 10^{-7}\left|{g_{\Sigma_b^+\to\ov B^0p}\over 5}\right|^2,
 \en
where the upper limit corresponds to $\Gamma_{\rm PV}/\Gamma_{\rm
PC}=1$. Therefore, the above result is consistent with the
experimental limit $1.6\times 10^{-6}$ \cite{Belle}.

We have chosen $|g_{\Sigma_b^+\to\ov B^0p}|=5$ as a benchmarked
value for the strong coupling for two reasons. First, a
calculation based on the $^3P_0$ quark-pair-creation model yields
a value of $6\sim 10$ for this coupling \cite{Jarfi}. Second, we
have computed the decay $B^-\to\Lambda_c\bar p\pi^-$ in
\cite{CYBbaryon}. A fit to the measured branching ratio for this
mode implies a strong coupling $g_{\Lambda_b\to B^-p}$ with the
strength in the vicinity of order 16. Using the relation
$|g_{\Lambda_b\to B^-p}|=3\sqrt{3/2}\,|g_{\Sigma_b^+\to \ov B^0p
}|$ derived from the $^3P_0$ quark-pair-creation model, it follows
that $|g_{\Sigma_b^+\to\ov B^0p}|\sim 4.4$, which is close to the
above-mentioned model estimate.

Note that a similar pole model calculation by Jarfi {\it et al.}
\cite{Jarfi} yields a branching ratio of order $7.0\times 10^{-5}$
after scaling their original result (see Table I of \cite{Jarfi})
to $|V_{ub}/V_{cb}|=0.085$ and to the current world average of $B$
lifetimes \cite{PDG}. Since $\Gamma_{\rm PV}/\Gamma_{\rm PC}=0.79$
is obtained by the same authors, and a strong coupling
$|g_{\Sigma_b^+\to\ov B^0p}|=10$ is used by them, it follows that
 \be
 \B(\ov B^0\to p\bar p)_{\rm PC}^{\rm H.O.}=1.0\times
 10^{-6}\left|{g_{\Sigma_b^+\to\ov B^0p}\over 5}\right|^2
 \en
is predicted by Jarfi {\it et al.} \cite{Jarfi} using the harmonic
oscillator wave functions for baryons. Evidently, the estimate of
the PC matrix element $a_{\Sigma_b^+p}$ in the harmonic oscillator
model is about three times as big as the one calculated in the bag
model.\footnote{It is not clear to us how to make a direct
comparison of our result for $a_{\Sigma_b^+p}$, which has a
dimension of mass, with the numerical value of $a_{\Sigma_b^+p}$
shown in Table II of \cite{Jarfi} which seems to be
dimensionless.}

 \vskip 0.6 cm
\centerline{\underline{2.~~$\ov B^0\to n\,\bar n$, $B^-\to n\bar
p$}}
 \vskip 0.4cm
The relevant intermediate states in the pole diagrams for the
decays $\ov B^0\to n\,\bar n$ and $B^-\to n\bar p$ are
$\Lambda_b^{(*)}$ and $\Sigma_b^{0(*)}$. Consider the former decay
first. The PV and PC wave amplitudes read
 \be
 A=-{g_{\Sigma_b^{0*}\to\ov
 B^0n}\,b_{\Sigma_b^{0*} n}\over m_n-m_{\Sigma_b^*}}-{g_{\Lambda_b^*\to\ov
 B^0n}\,b_{\Lambda_b^* n}\over m_n-m_{\Lambda_b^*}}, \qquad B=\,{g_{\Sigma_b^0\to\ov
 B^0n}\,a_{\Sigma_b^0 n}\over m_n-m_{\Sigma_b}}+{g_{\Lambda_b\to\ov
 B^0n}\,a_{\Lambda_b n}\over m_n-m_{\Lambda_b}}.
 \en
Applying the bag model leads to the PC matrix elements
 \be \label{nSigmab}
 \la n|O_1^{\rm PC}|\Sigma_b^0\ra=3\sqrt{2}X(4\pi), \qquad\quad
 \la n|O_1^{\rm PC}|\Lambda_b\ra=\sqrt{6}X(4\pi).
 \en
For strong couplings,  the $^3P_0$ quark-pair-creation model
implies \cite{Yaouanc}
 \be
{g_{\Lambda_b\to\ov B^0n}\over g_{\Sigma_b^0\to\ov
 B^0n}}={\la \Phi_{n^\up}(124)\Phi_{\ov B^0}(35)|\Phi_{\Lambda_b^{\up}}(123)\Phi_{\rm
vac}(45)\ra\over \la \Phi_{n^\up}(124)\Phi_{\ov
B^0}(35)|\Phi_{\Sigma_b^{0\up}}(123)\Phi_{\rm vac}(45)\ra },
 \en
where the $\Phi$'s are the spin-flavor wave functions and the
vacuum wave function has the expression
 \be
 \Phi_{\rm vac}={1\over\sqrt{3}}(u\bar u+d\bar d+s\bar s)\otimes
 {1\over\sqrt{2}}(\up\dw+\dw\up).
 \en
Using the baryon wave functions given in Eq. (\ref{spin-flavor})
and the $B$ meson wave function
 \be
\Phi_{\ov B^0}=b\,\bar d\otimes {1\over\sqrt{2}}(\up\dw-\dw\up),
 \en
we obtain
 \be \label{grel3}
 g_{\Lambda_b\to\ov B^0n}=-3\sqrt{3}\,g_{\Sigma_b^0\to\ov
 B^0n}.
 \en
Consequently,
 \be \label{Bnn}
 B(\ov B^0\to n\bar n)=-g_{\Sigma_b^0\to\ov B^0n}\left( {3\sqrt{3}\,a_{\Lambda_b n}\over
 m_n-m_{\Lambda_b}}-{a_{\Sigma_b^0 n}\over
 m_n-m_{\Sigma_b}}\right).
 \en
Likewise, for $B^-\to n\bar p$ we have
 \be \label{Bnp}
 B(B^-\to n\bar p)=g_{\Sigma_b^0\to\ov B^0n}\left( {3\sqrt{3}\,a_{\Lambda_b n}\over
 m_p-m_{\Lambda_b}}+{a_{\Sigma_b^0 n}\over
 m_p-m_{\Sigma_b}}\right),
 \en
where use has been made of
 \be \label{grel1}
 g_{\Lambda_b\to B^-p}=3\sqrt{3}\,g_{\Sigma_b^0\to\ov
 B^0n}, \qquad\quad g_{\Sigma_b^0\to B^-p}=g_{\Sigma_b^0\to\ov
 B^0n}.
 \en
Using the relations
 \be \label{grel2}
 g_{\Sigma_b^0\to\ov B^0n}=-{1\over\sqrt{2}}g_{\Sigma_b^+\to\ov
 B^0p},
 \en
and
 \be
 a_{\Sigma_b^0n}=-{1\over\sqrt{2}}\,a_{\Sigma_b^+p}
 \en
derived from Eqs. (\ref{pSigmab}) and (\ref{nSigmab}), we find
that $\ov B\to N\ov N$ amplitudes satisfy the $\Delta I=1/2$
relation \cite{Korner,Jarfi}
 \be  \label{isospin}
 \A(\ov B^0\to p\bar p)-\A(\ov B^0\to n\bar n)=\A(B^-\to n\bar p).
 \en
As mentioned before, this $\Delta I={1\over 2}$ relation arises
because the weak operator $O_1-O_2$ has isospin $I={1\over 2}$.

From Eqs. (\ref{nSigmab}), (\ref{Bnn}) and (\ref{Bnp}), it is
evident that $B^-\to n\bar p$ has a larger rate than $\ov B^0\to
n\bar n$. In contrast, the QCD sum rule analysis in
\cite{Chernyak} predicts that $\Gamma(\ov B^0\to p\bar
p)>\Gamma(B^-\to n\bar p)>\Gamma(\ov B^0\to n\bar n)$. Moreover,
as pointed out in \cite{Korner,Jarfi}, the decay $B^-\to n\bar p$
is purely parity-conserving, namely,  its parity-violating
amplitude vanishes provided that the $q\bar q$ pair is created
from the vacuum. As pointed out by K\"orner \cite{Korner}, if the
quark pair is created perturbatively via one gluon exchange with
one-gluon quantum number ($^3S_1$ model), the neutron in $B^-\to
n\bar p$ will have a positive longitudinal polarization.
Therefore, a polarization measurement of the neutron by studying
its subsequent weak decay can be used to test the $^3P_0$ and
$^3S_1$ quark-pair-creation models.

We are ready to compute branching ratios and obtain
 \be
 \B(\ov B^0\to n\bar n)_{\rm PC} &=& 1.2\times
 10^{-7}\left|{g_{\Sigma_b^+\to\ov B^0p}\over 5}\right|^2, \non \\
 \B(B^-\to n\bar p) &=& 5.0\times
 10^{-7}\left|{g_{\Sigma_b^+\to\ov B^0p}\over 5}\right|^2.
 \en

\vskip 0.6cm  \centerline{\underline{3.~~$\ov
B^0\to\Lambda\bar\Lambda$}} \vskip 0.4cm

Let us consider the PC amplitude of $\ov B^0\to
\Lambda\bar\Lambda$. In the pole model it receives pole
contributions from the anti-triplet $\Xi_b^{0}$ and sextet
$\Xi_b^{0'}$
 \be
 B=\,{g_{\Xi_b^{0}\to\ov B^0\Lambda}\,a_{\Xi_b^{0}\Lambda}\over
 m_\Lambda-m_{\Xi_b}}+{g_{\Xi_b^{'0}\to\ov B^0\Lambda}\,a_{\Xi_b^{'0}\Lambda}\over
 m_\Lambda-m_{\Xi_b'}}.
 \en
Using the wave functions given in Appendix A, we obtain
 \be
 \la \Lambda|O_1^{\rm PC}|\Xi_b^{0}\ra &=& -2X(4\pi), \non \\
 \la \Lambda|O_1^{\rm PC}|\Xi_b^{'0}\ra &=& -2\sqrt{3}X(4\pi),
 \en
for PC matrix elements, and
 \be
 g_{\Xi_b^{0}\to\ov B^0\Lambda}=-\sqrt{3}\,g_{\Xi_b^{'0}\to\ov B^0\Lambda}
 \en
for strong couplings. Then it is clear that the PC amplitude
vanishes as the mass difference between $\Xi_b$ and $\Xi_b'$ is
negligible. That is, this decay is purely parity violating in the
$^3P_0$ quark-pair-creation model as noticed by K\"orner
\cite{Korner} and Jarfi {\it et al.} \cite{Jarfi} some time ago.
As noted in passing, we will not compute the PV amplitude within
the framework of the bag model.

\vskip 0.6cm  \centerline{\underline{4.~~$B^-\to p
\bar\Delta^{--},~n\bar\Delta^-$, $\ov B^0\to p
\bar\Delta^-,~n\bar\Delta^0$}} \vskip 0.4cm

The relevant pole diagram consists of the intermediate states
$\Sigma_b^{+(*)}$ for $p\bar\Delta^{--}$, $p\bar\Delta^-$ modes
($\bar\Delta^{--}$ being the antiparticle of $\Delta^{++}$ and
likewise for other $\bar\Delta$ particles) and $\Sigma_b^{0(*)}$
as well as $\Lambda_b^{(*)}$ for $n\bar\Delta^-$, $n\bar\Delta^0$
final states. However,  it is straightforward to show that, in the
$^3P_0$ quark-pair-creation model, the strong coupling for
$\Lambda_b\to N\bar\Delta$ vanishes and hence the $\Lambda_b$ pole
makes no contribution. Moreover, the parity-violating part
vanishes in the same quark-pair-creation model
\cite{Korner,Jarfi}. Therefore,
 \be
&&  D(B^-\to p\bar\Delta^{--})=\,{g_{\Sigma_b^+\to
B^-\Delta^{++}}\,
 a_{\Sigma_b^+ p}\over m_p-m_{\Sigma_b}}, \qquad\quad  D(B^-\to n\bar\Delta^{-})
 =\,{g_{\Sigma_b^0\to B^-\Delta^{+}}\,a_{\Sigma_b^0 n}\over
 m_n-m_{\Sigma_b}},   \non \\
&& D(\ov B^0\to p\bar\Delta^{-})=\,{g_{\Sigma_b^+\to \ov
 B^0\Delta^{+}}\,a_{\Sigma_b^+ p}\over m_p-m_{\Sigma_b}}, \qquad\quad
 D(\ov B^0\to n\bar\Delta^{0})
 =\,{g_{\Sigma_b^0\to \ov B^0\Delta^{0}}\,a_{\Sigma_b^0 n}\over
 m_n-m_{\Sigma_b}},
 \en
where the PC matrix elements $a_{\Sigma_b^+p}$ and $a_{\Sigma_b^0
n}$ have been evaluated before. The relative strong couplings are
 \be \label{gDelta}
 g_{\Sigma_b^+\to B^-\Delta^{++}} &=& -\sqrt{3}\,g_{\Sigma_b^+\to \ov B^0\Delta^{+}}
 =-\sqrt{3/2}\, g_{\Sigma_b^0\to \ov B^0\Delta^{0}} \non \\
 &=& \sqrt{3/2}\, g_{\Sigma_b^0\to B^-\Delta^{+}}= 2\sqrt{6}\,g_{\Sigma_b^+\to\ov
 B^0p}\,.
 \en
This together with the baryon matrix elements (\ref{pSigmab}) and
(\ref{nSigmab}) leads to the relation
 \be \label{NDelta}
 \Gamma(B^-\to p\bar\Delta^{--})=3\Gamma(B^-\to n\bar\Delta^{-})=
 3\Gamma(\ov B^0\to p\bar\Delta^{-})=3\Gamma
 (\ov B^0\to n\bar\Delta^{0}),
 \en
as first pointed out by Jarfi {\it et al.} \cite{Jarfi}. In the
diquark model of \cite{Ball}, $n\bar\Delta^-$ has a rate different
from $p\bar\Delta^-$ and $n\bar\Delta^0$. Hence, experimentally it
is important to test the relation (\ref{NDelta}).

If we apply Eq. (\ref{gDelta}) and use $|g_{\Sigma_b^+\to\ov
B^0p}|=5$, we will obtain $|g_{\Sigma_b^+\to B^-\Delta^{++}}|=24$
and $\B(B^-\to p\bar\Delta^{--})=5.8\times 10^{-6}$. Because of
the strong decay $\bar\Delta^{--}\to\bar p\pi^-$, the resonant
contribution from $\bar\Delta^{--}$ to the branching ratio of
$p\bar p\pi^-$ would be $6\times 10^{-6}$. This already exceeds
the recent Belle measurement $\B(B^-\to p\bar
p\pi^-)=(1.9^{+1.0}_{-0.9}\pm0.3)\times 10^{-6}$ or the upper
limit of  $\B(B^-\to p\bar p\pi^-)<3.7\times 10^{-6}$
\cite{Bellebaryon}. Therefore, the coupling of the $\Delta$ to the
$B$ meson and the octet baryon is smaller than what is expected
from Eq. (\ref{gDelta}) probably due to the different
off-shellness of $\Delta$. Recall that the parity-conserving
transition to the $\Delta$ corresponds to a $L=2$ partial wave.
Therefore, the off-shell suppression on the three-point coupling
of $\Sigma_b\to B\Delta$ is likely to be different from that of
$\Lambda_b\to BN$. For definiteness, we will choose
$|g_{\Sigma_b^+\to B^-\Delta^{++}}|=12$ and obtain
 \be \label{pDelta}
 \B(B^-\to p\bar\Delta^{--})=1.4\times 10^{-6}\left|{g_{\Sigma_b^+\to B^-\Delta^{++}}
 \over 12}\right|^2.
 \en
Thus this charmless decay $B^-\to p\bar\Delta^{--}$ can have a
large branching ratio of order $10^{-6}$ owing to the large
coupling constant $g_{\Sigma_b^+\to B^-\Delta^{++}}$. In sharp
contrast, this mode is predicted to be only at the level of
$3\times 10^{-7}$ in the QCD sum rule analysis \cite{Chernyak}
(see also Table II). The branching ratios of other modes can be
calculated using Eq. (\ref{NDelta}) and are shown in Table II.
Experimentally, the decay $B^-\to p\bar\Delta^{--}$ should be
readily accessible by $B$ factories BaBar and Belle.

\subsection{Penguin-dominated two-body decays}

\vskip .6cm \centerline{ \underline{1.~~$B^-\to\Lambda\bar p$,
$\ov B^0\to\Lambda\bar n$}} \vskip 0.4cm

This decay receives internal $W$-emission and $b\to s$ penguin
contributions [see Fig. 1(a)]. As we shall see below, it is a
penguin-dominated mode.  The pole diagram for $B^-\to\Lambda\bar
p$ consists of the intermediate states $\Lambda_b^{0(*)}$ and
$\Sigma_b^{0(*)}$
 \be
 A = -{ g_{\Lambda_b^*\to B^-p}\,b_{\Lambda_b^*\Lambda}\over
 m_\Lambda-m_{\Lambda_b^*}}-{ g_{\Sigma_b^{0*}\to B^-p}\,b_{\Sigma_b^{0*}\Lambda}\over
 m_\Lambda-m_{\Sigma_b^*}}, \qquad
B = { g_{\Lambda_b\to B^-p}\,a_{\Lambda_b\Lambda}\over
 m_\Lambda-m_{\Lambda_b}}+{ g_{\Sigma_b^0\to B^-p}\,a_{\Sigma_b^0\Lambda}\over
 m_\Lambda-m_{\Sigma_b}}.
 \en
To evaluate the hadronic matrix elements, we notice that the
combinations of the operators $O_{2i+1}+O_{2i+2}$ $(i=0,\cdots,4)$
are symmetric in color indices and hence they cannot contribute to
the baryon-baryon matrix element. From this we can write the PC
matrix element $a_{\Lambda_b\Lambda}$ as
 \be \label{Lambm.e.1}
 a_{\Lambda_b\Lambda} &=&
 {G_F\over\sqrt{2}}\Bigg\{V_{ub}V_{us}^*(c_1^{\rm eff}-c_2^{\rm
 eff})\la \Lambda|O_1^{\rm PC}|\Lambda_b\ra -V_{tb}V_{ts}^*\Big[(c_3^{\rm eff}-c_4^{\rm
 eff})\la \Lambda|O_3^{\rm PC}|\Lambda_b\ra \non \\
 &+& (c_5^{\rm eff}-c_6^{\rm eff})\la \Lambda|O_5^{\rm PC}|\Lambda_b\ra+(c_7^{\rm eff}-c_8^{\rm
 eff})\la \Lambda|O_7^{\rm PC}|\Lambda_b\ra +(c_9^{\rm eff}-c_{10}^{\rm
 eff})\la \Lambda|O_9^{\rm PC}|\Lambda_b\ra\Big]\Bigg\}.
 \en
Since the bag model implies
 \be
 \la\Lambda|(\bar s b)_\vma(\bar dd)_{_{V\pm A}}|\Lambda_b\ra_{\rm PC}=
 \la\Lambda|(\bar s b)_\vma(\bar uu)_{_{V\pm A}}|\Lambda_b\ra_{\rm
 PC},
 \en
the baryon matrix elements of $O_3$ and $O_9$ can be related to
$O_1$, while matrix element of $O_7$ is related to $O_5$, for
example,
 \be
 \la\Lambda|O_3^{\rm PC}|\Lambda_b\ra &=& \la\Lambda|(\bar sb)_\vma[(\bar uu)_\vma
 +(\bar dd)_\vma]|\Lambda_b\ra_{\rm PC}=-2\la \Lambda
 |O_1^{\rm PC}|\Lambda_b\ra.
 \en
Hence, Eq. (\ref{Lambm.e.1}) can be recast as
 \be \label{Lambm.e.2}
 a_{\Lambda_b\Lambda} &=&
 {G_F\over\sqrt{2}}\Bigg\{\Big[V_{ub}V_{us}^*(c_1^{\rm eff}-c_2^{\rm
 eff}) -V_{tb}V_{ts}^*(-2c_3^{\rm eff}+2c_4^{\rm
 eff}-{1\over 2}c_9^{\rm eff}+{1\over 2}c_{10}^{\rm eff})\Big]\la
 \Lambda|O_1^{\rm PC}|\Lambda_b\ra \non \\
 && -V_{tb}V_{ts}^*(c_5^{\rm eff}-c_6^{\rm eff}+{1\over 2}c_7^{\rm eff}-{1\over 2}c_8^{\rm
 eff})\la \Lambda|O_5^{\rm PC}|\Lambda_b\ra \Bigg\}.
 \en
Likewise, the relation
 \be
 \la\Lambda|(\bar s b)_\vma(\bar dd)_{_{V\pm A}}|\Sigma_b^0\ra_{\rm PC}=-
 \la\Lambda|(\bar s b)_\vma(\bar uu)_{_{V\pm
 A}}|\Sigma_b^0\ra_{\rm PC}
 \en
implied by the bag model leads to
 \be
 a_{\Sigma_b^0\Lambda} &=&
 {G_F\over\sqrt{2}}\Bigg\{\Big[V_{ub}V_{us}^*(c_1^{\rm eff}-c_2^{\rm
 eff}) -V_{tb}V_{ts}^*(-{3\over 2}c_9^{\rm eff}+{3\over 2}c_{10}^{\rm eff})\Big]
 \la \Lambda|O_1^{\rm PC}|\Sigma_b^0\ra \non \\
 && -V_{tb}V_{ts}^*(c_7^{\rm eff}-c_8^{\rm
 eff})\la \Lambda|O_7^{\rm PC}|\Sigma_b^0\ra \Bigg\}.
 \en
Therefore, the PC matrix element for $\Sigma_b^0-\Lambda$ weak
transition does not receive QCD penguin contributions.

Applying Eqs. (\ref{mevma}) and (\ref{mevpa}) we obtain
 \be
\la \Lambda|O_1^{\rm PC}|\Lambda_b\ra &=&{4\over 3}X_1(4\pi), \non \\
\la \Lambda|O_5^{\rm PC}|\Lambda_b\ra &=&{4\over
3}(2Y_1+2Y_2-Y_1'+Y_2')(4\pi),  \\
\la \Lambda|O_1^{\rm PC}|\Sigma_b^0\ra &=&-{1\over \sqrt{3}}(X_1+3X_2)(4\pi), \non \\
\la \Lambda|O_7^{\rm PC}|\Sigma_b^0\ra &=&-{\sqrt{3}\over
2}[Y_1+Y_2-2(Y_1'-Y_2')](4\pi),  \non
 \en
in the bag model, where
 \be
 X_1 &=& \int^R_0
 r^2dr[u_s(r)v_u(r)-v_s(r)u_u(r)][u_u(r)v_b(r)-v_u(r)u_b(r)], \non
 \\
 X_2 &=& \int^R_0
 r^2dr[u_s(r)u_u(r)+v_s(r)v_u(r)][u_u(r)u_b(r)+v_u(r)v_b(r)], \non
 \\
 Y_1 &=& \int^R_0
 r^2dr[u_s(r)v_b(r)-v_s(r)u_b(r)][u_u(r)v_u(r)-v_u(r)u_u(r)], \non
 \\
 Y_1' &=& \int^R_0
 r^2dr[u_s(r)v_b(r)+v_s(r)u_b(r)][u_u(r)v_u(r)+v_u(r)u_u(r)],
 \\
 Y_2 &=& \int^R_0
 r^2dr[u_s(r)u_b(r)+v_s(r)v_b(r)][u_u(r)u_u(r)+v_u(r)v_u(r)], \non
 \\
 Y_2' &=& \int^R_0
 r^2dr[u_s(r)u_b(r)-v_s(r)v_b(r)][u_u(r)u_u(r)-v_u(r)v_u(r)], \non
 \en
are four-quark overlap bag integrals. Finally we arrive at
 \be
 B(B^-\to\Lambda\bar p)= -{1\over \sqrt{2}}g_{\Sigma_b^+\to \ov B^0p}
 \left( {3\sqrt{3}\,a_{\Lambda_b\Lambda}\over
 m_\Lambda-m_{\Lambda_b}}+{a_{\Sigma_b^0\Lambda}\over
 m_\Lambda-m_{\Sigma_b}}\right),
 \en
where use has been made of Eqs. (\ref{grel1}) and (\ref{grel2}).

The bag integrals have the values
 \be
&& X_1=-4.6\times 10^{-6}\,{\rm GeV}^3, \qquad X_2=1.7\times
10^{-4}\,{\rm GeV}^3, \qquad
 Y_1=0,  \non \\
&& Y_1'=4.5\times 10^{-5}\,{\rm GeV}^3, \qquad Y_2=1.7\times
10^{-4}\,{\rm GeV}^3, \qquad Y_2'=1.2\times 10^{-4}\,{\rm GeV}^3.
 \en
It is easy to check that $a_{\Lambda_b\Lambda}$ and hence the
decay is penguin dominated.  For the branching ratio we find
 \be \label{BRLambdap}
 \B(B^-\to\Lambda \bar p)_{\rm PC}=2.2\times 10^{-7}\left|{g_{\Sigma_b^+\to
 \ov B^0p}\over 5}\right|^2.
 \en
For $\ov B^0\to\Lambda \bar n$, it has the same rate as
$B^-\to\Lambda \bar p$.

\vskip 0.6cm \centerline{ \underline{2.~~$\ov B^0\to\Sigma^+\bar
p$}} \vskip 0.4cm

We consider the pole diagram with the intermediate states
$\Sigma_b^{+(*)}$
 \be
 A=-{g_{\Sigma_b^{+*}\to \ov B^0p}b_{\Sigma_b^*p}\over
 m_\Sigma-m_{\Sigma_b^{*}} }, \qquad\qquad B=\,{g_{\Sigma_b^+\to\ov B^0p}
 a_{\Sigma_b^+\Sigma}\over  m_\Sigma-m_{\Sigma_b^+}}.
 \en
The PC weak matrix element for $\Sigma_b^+-\Sigma^+$ transition
reads
 \be
 a_{\Sigma_b^+\Sigma^+} &=&
 {G_F\over\sqrt{2}}\Bigg\{\Big[V_{ub}V_{us}^*(c_1^{\rm eff}-c_2^{\rm
 eff}) -V_{tb}V_{ts}^*(-c_3^{\rm eff}+c_4^{\rm
 eff}-c_9^{\rm eff}+c_{10}^{\rm eff})\Big]\la \Sigma^+|O_1^{\rm PC}|\Sigma_b^+\ra \non \\
 && -V_{tb}V_{ts}^*(c_5^{\rm eff}-c_6^{\rm eff}+c_7^{\rm eff}-c_8^{\rm
 eff})\la \Sigma^+|O_5^{\rm PC}|\Sigma_b^+\ra \Bigg\}.
 \en
In the bag model,
 \be
\la \Sigma^+|O_1^{\rm PC}|\Sigma_b^+\ra &=&{2\over 3}(X_1-9X_2)(4\pi), \non \\
\la \Sigma^+|O_5^{\rm PC}|\Sigma_b^+\ra &=&{2\over
3}[Y_1+Y_2+4(Y_1'-Y_2')](4\pi).
 \en
We obtain numerically
 \be
 \B(\ov B^0\to\Sigma^+ \bar p)_{\rm PC}=1.8\times 10^{-8}\left|{g_{\Sigma_b^+\to
 \ov B^0 p}\over 5}\right|^2.
 \en
Note that the branching ratio is predicted to be $5\times 10^{-6}$
in the QCD sum rule analysis of \cite{Chernyak}, which is larger
than our result by two orders of magnitude (see Table II).

\vskip 0.6cm \centerline{ \underline{3.~~$B^-\to\Sigma^0\bar p$}}
\vskip 0.4cm

The intermediate low-lying pole states for this decay are
$\Lambda_b^{0(*)}$ and $\Sigma_b^{0(*)}$
 \be
 A = -{ g_{\Lambda_b^*\to B^-p}\,b_{\Lambda_b^*\Sigma^0}\over
 m_\Sigma-m_{\Lambda_b^*}}-{ g_{\Sigma_b^{0*}\to B^-p}\,b_{\Sigma_b^{0*}\Sigma^0}\over
 m_\Sigma-m_{\Sigma_b^*}}, \qquad
 B = { g_{\Lambda_b\to B^-p}\,a_{\Lambda_b\Sigma^0}\over
 m_\Sigma-m_{\Lambda_b}}+{ g_{\Sigma_b^0\to B^-p}\,a_{\Sigma_b^0\Sigma^0}\over
 m_\Sigma-m_{\Sigma_b}}.
 \en
The PC matrix elements are given by
 \be
 a_{\Sigma_b^0\Sigma^0} &=&
 {G_F\over\sqrt{2}}\Bigg\{\Big[V_{ub}V_{us}^*(c_1^{\rm eff}-c_2^{\rm
 eff}) -V_{tb}V_{ts}^*(-2c_3^{\rm eff}+2c_4^{\rm
 eff}-{1\over 2}c_9^{\rm eff}+{1\over 2}c_{10}^{\rm eff})\Big]\la \Sigma^0|O_1^{\rm PC}|\Sigma_b^0\ra \non \\
 && -V_{tb}V_{ts}^*(c_5^{\rm eff}-c_6^{\rm eff}+{1\over 2}c_7^{\rm eff}-{1\over 2}c_8^{\rm
 eff})\la \Sigma^0|O_5^{\rm PC}|\Sigma_b^0\ra \Bigg\},
 \en
and
 \be
 a_{\Lambda_b^0\Sigma^0} &=&
 {G_F\over\sqrt{2}}\Bigg\{\Big[V_{ub}V_{us}^*(c_1^{\rm eff}-c_2^{\rm
 eff}) -V_{tb}V_{ts}^*(-{3\over 2}c_9^{\rm eff}+{3\over 2}c_{10}^{\rm eff})\Big]
 \la \Sigma^0|O_1^{\rm PC}|\Lambda_b\ra \non \\
 && -V_{tb}V_{ts}^*(c_7^{\rm eff}-c_8^{\rm
 eff})\la \Sigma^0|O_7^{\rm PC}|\Lambda_b\ra \Bigg\},
 \en
where in the bag model
 \be
\la \Sigma^0|O_1^{\rm PC}|\Lambda_b\ra &=&-{1\over \sqrt{3}}(X_1+3X_2)(4\pi), \non \\
\la \Sigma^0|O_7^{\rm PC}|\Lambda_b\ra &=&-{\sqrt{3}\over
2}[Y_1+Y_2-2(Y_1'-Y_2')](4\pi),  \\
\la \Sigma^0|O_1^{\rm PC}|\Sigma_b^0\ra &=&{1\over 3}(X_1-9X_2)(4\pi), \non \\
\la \Sigma^0|O_5^{\rm PC}|\Sigma_b^0\ra &=&{2\over
3}[Y_1+Y_2+4(Y_1'-Y_2')](4\pi). \non
 \en
Hence,
 \be
 B= -{1\over \sqrt{2}}g_{\Sigma_b^+\to \ov B^0p}\left( {3\sqrt{3}\,a_{\Lambda_b\Sigma^0}\over
 m_\Sigma-m_{\Lambda_b}}+{a_{\Sigma_b^0\Sigma^0}\over
 m_\Sigma-m_{\Sigma_b}}\right),
 \en
where use of Eqs. (\ref{grel1}) and (\ref{grel2}) for strong
couplings has been made. We obtain
 \be
 \B(B^-\to\Sigma^0 \bar p)_{\rm PC}=5.8\times 10^{-8}\left|{g_{\Sigma_b^+\to
 \ov B^0 p}\over 5}\right|^2.
 \en
Again, the QCD sum rule prediction for this mode is much higher
\cite{Chernyak}.

\vskip 0.6cm \centerline{
\underline{4.~~$B^-\to\Sigma^+\bar\Delta^{--},~\Sigma^-\bar\Delta^0$,
$\ov B^0\to\Sigma^+\bar\Delta^-$}} \vskip 0.4cm

As stated before, the decays
$B^-\to\Sigma^+\bar\Delta^{--},~\Sigma^-\bar\Delta^0$, $\ov
B^0\to\Sigma^+\bar\Delta^-$ only receive parity-conserving
contributions \cite{Korner,Jarfi} so that
 \be
 && D(B^-\to\Sigma^+\bar\Delta^{--})=\,{g_{\Sigma_b^+\to
B^-\Delta^{++}}\, a_{\Sigma_b^+ \Sigma^+}\over
m_\Sigma-m_{\Sigma_b}}, \non  \\
 && D(\ov
B^0\to\Sigma^+\bar\Delta^{-})=\,{g_{\Sigma_b^+\to \ov
 B^0\Delta^{+}}\,a_{\Sigma_b^+ \Sigma^+}\over m_\Sigma-m_{\Sigma_b}},
  \\
&& D(B^-\to\Sigma^-\bar\Delta^{0})
 =\, {g_{\Sigma_b^-\to B^-\Delta^{0}}\,a_{\Sigma_b^- \Sigma^-}\over
 m_\Sigma-m_{\Sigma_b}}. \non
 \en
The PC matrix element $a_{\Sigma_b^+\Sigma^+}$ has been evaluated
before and $a_{\Sigma^-_b\Sigma^-}=a_{\Sigma_b^+\Sigma^+}$. For
strong couplings we get
 \be
 g_{\Sigma_b^+\to\Delta^{++}B^-}=-\sqrt{3}\,g_{\Sigma_b^+\to\Delta^+\ov
 B^0}=\sqrt{3}\,g_{\Sigma_b^-\to\Delta^0
 B^-}=2\sqrt{6}\,g_{\Sigma_b^+\to p\ov B^0},
 \en
in the $^3P_0$ model. Collecting all the results gives
 \be
 \B(B^-\to\Sigma^+\bar\Delta^{--}) &=& 2.0\times 10^{-7}\left|{g_{\Sigma_b^+\to
 \Delta^{++}B^-}\over 12}\right|^2, \non \\
 \B(\ov B^0\to\Sigma^+\bar\Delta^{-}) &=& 6.3\times 10^{-8}\left|{g_{\Sigma_b^+\to
 \Delta^{++}B^-}\over 12}\right|^2, \non \\
 \B(B^-\to\Sigma^-\bar\Delta^{0}) &=& 8.7\times 10^{-8}\left|{g_{\Sigma_b^+\to
 \Delta^{++}B^-}\over 12}\right|^2,
 \en
where we have followed the discussion of $\ov B\to N\bar\Delta$ to
choose the coupling $|g_{\Sigma_b^+\to
 \Delta^{++}B^-}|=12$ as a benchmarked value.

\subsection{Comparison with other models}
As discussed in passing, though we adopt the same pole-model
framework as Jarfi {\it et al.} \cite{Jarfi} for describing
two-body baryonic $B$ decays, a crucial difference is that weak
baryon matrix elements are evaluated by Jarfi {\it et al.} at
large momentum transfer and strong couplings at small transfer,
whereas the weak transition is computed at zero transfer and
strong couplings at large momentum transfer in our case. In
general, the difference in numerical results shown in Table II
comes mainly from the fact that we use the bag model rather than
the harmonic oscillator model to evaluate weak baryon transitions.

In the following we compare our results with the diquark model by
Ball and Dosch \cite{Ball} and the QCD sum rule analysis by
Chernyak and Zhitnitsky \cite{Chernyak} (see also Table II). For
$\ov B\to N\bar N$ decays, the diquark model has one unique
prediction, namely, there is no $B^-\to n\bar p$ decay, while
$p\bar p$ and $n\bar n$ final states have the same rates. In
contrast, the sum rule approach predicts that $\Gamma(\ov B^0\to
p\bar p)>\Gamma(B^-\to n\bar p)>\Gamma(\ov B^0\to n\bar n)$ (see
Table II), while in our case $\Gamma(B^-\to n\bar p)>\Gamma(\ov
B^0\to p\bar p)\approx \Gamma(\ov B^0\to n\bar n)$. Therefore, a
measurement of the relative rates of $\ov B\to N\bar N$
(especially $B^-\to n\bar p$) will serve to test the three models.

As for the tree-dominated modes $\ov B\to N\bar\Delta$, they are
suppressed in the diquark model because the operators $O_1$ and
$O_2$ can only generate scalar diquarks whereas the decuplet
baryons are made of a vector diquark and a quark. Likewise, they
are also suppressed in the sum rule analysis. In sharp contrast,
these modes have sizable branching ratios in the pole model,
namely, $\Gamma(\ov B\to N\bar\Delta)>\Gamma(\ov B\to N\bar N)$,
owing to the large coupling of the intermediate state $\Sigma_b$
with the $B$ meson and the $\Delta$ resonance.

The penguin-dominated decays have smaller rates than $\ov B\to
p\bar p$ in the diquark model as the penguin operators are not
included in the original calculations by Ball and Dosch (the
effect of the penguin operators in this model was recently
discussed in \cite{CH}). In contrast, the sum rule approach
predicts branching ratios of order $(2-6)\times 10^{-6}$ for $\ov
B\to\Lambda\bar p,~\Sigma\bar p,~\Sigma\bar\Delta$. In our work,
the decay rates of penguin-dominated decays are in general small.

In short, measurements of the relative rates of $\ov B\to N\bar
N,~\Sigma\bar p,~\Sigma\bar\Delta$ will suffice to differentiate
between above-mentioned three models.

\section{charmless three-body baryonic decays}
As noted in the Introduction, the study and search of the
three-body baryonic $B$ decay $\ov B\to\B_1\ov \B_2 M$ with $M$
being a meson are mainly motivated by the experimental observation
that $\B(B^-\to\Lambda_c\bar p\pi^-)>\B(\ov B^0\to\Lambda_c\bar
p)$ \cite{CLEOc} and $\B(B^-\to p\bar p K^-)>\B(\ov B^0\to p\bar
p)$ \cite{Bellebaryon}. Theoretically, it has been argued that the
emitted meson $M$ in the three-body final state carries away much
energies and the configuration is more favorable for baryon
production because of reduced energy release compared to the
latter \cite{CHT2}. Roughly speaking, the reason that the two-body
baryonic decay $B\to\B_1\ov\B_2$ is smaller than the mesonic
counterpart $B\to M_1\ov M_2$ stems from the fact that one needs
an additional quark pair production in the internal $W$-emission
diagram [Fig. 1(a)] and two $q\bar q$ pairs in weak annihilation
diagrams [Fig. 1(b)] in order to form a baryon-antibaryon pair. A
$q\bar q$ production is suppressed by either a strong coupling
when it is produced perturbatively via one gluon exchange or by
intrinsic softness of nonperturbative pair creation \cite{HS}. In
the three-body baryonic decay, the emission of the meson $M$ will
carry away energies in such a way that the invariant mass of
$\B_1\ov\B_2$ becomes smaller and hence it is relatively easier to
fragment into the baryon-antibaryon pair.

One can also understand the above feature more concretely by
studying the Dalitz plot. Due to the $V-A$ nature of the $b\to
ud\bar u$ process, the invariant mass of the diquark $ud$ peaks at
the highest possible values in a Dalitz plot for $b\to ud\bar d$
transition (see \cite{Buchalla} and footnote [91] in
\cite{Dunietz96}). If the $ud$ forms a nucleon, then the very
massive $udq$ objects will intend to form a highly excited baryon
state such as $\Delta$ and $N^*$ and will be seen as $N n\pi(n\geq
1)$ \cite{Dunietz}. This explains the non-observation of the $N\ov
N$ final states and why the three-body mode $N\ov N \pi(\rho)$ is
favored. Of course, this does not necessarily imply that the
three-body final state $\B_1\ov \B_2M$ always has a larger rate
than the two-body one $\B_1\ov\B_2$. In this section we will study
some leading charmless three-body baryonic $B$ decays and see
under what condition that the above argument holds.

\begin{figure}[p]
\hspace{2cm} \centerline{\psfig{figure=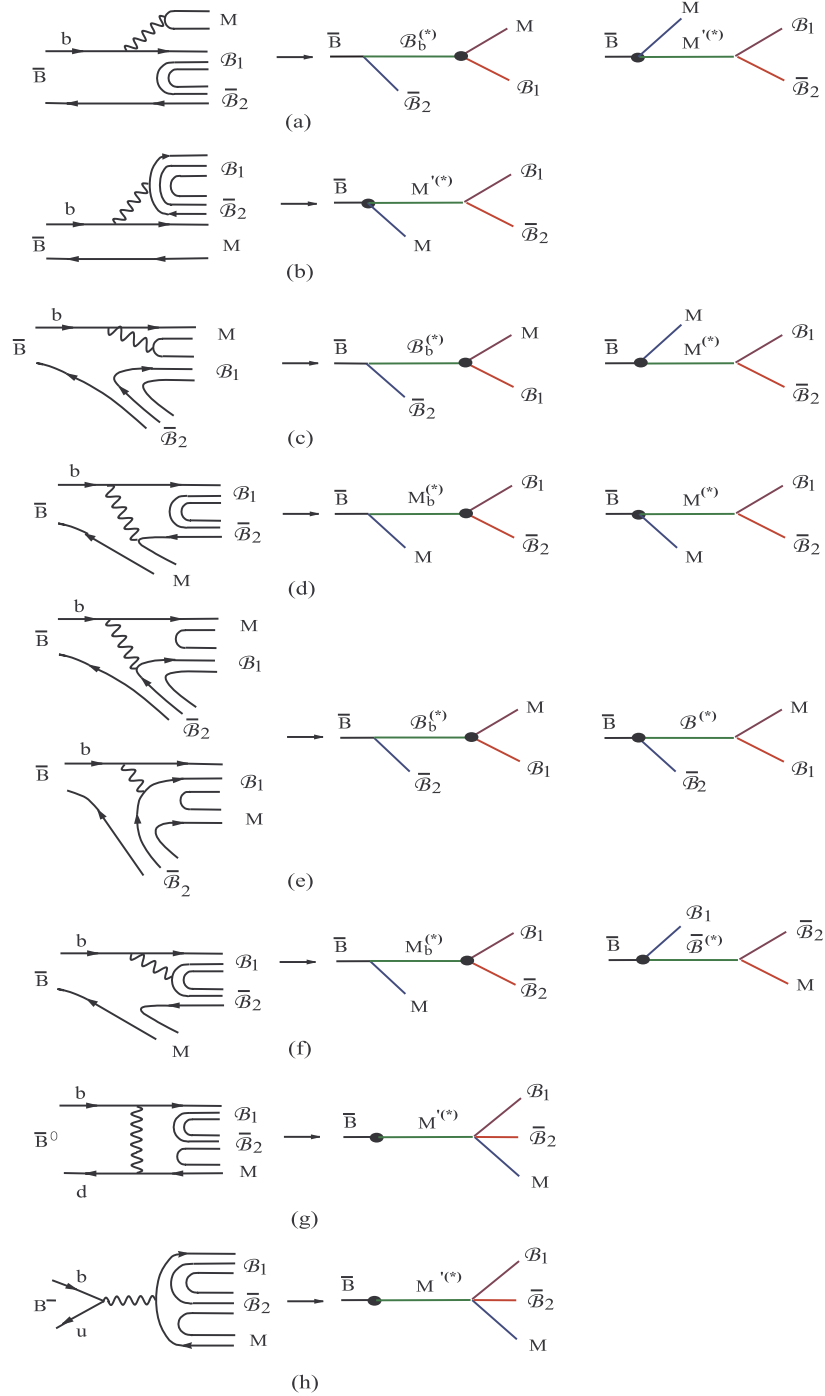,height=7.5in}}
\vspace{0.2cm}
    \caption{{\small Quark and pole diagrams for three-body baryonic $B$ decay
    $\ov B\to \B_1\ov \B_2 M$, where the symbol $\bullet$ denotes the weak vertex.
    Figs. 2(a) and  2(b) correspond to factorizable external $W$-emission
    contributions, Figs. 2(c) and 2(d) to factorizable internal
    $W$-emission, Figs. 2(e) and 2(f) to nonfactorizable internal
    $W$-emission, Fig. 2(g) to $W$-exchange and Fig. 2(h) to
    $W$-annihilation. Penguin contributions are obtained from Figs. 2(c)-2(g) by
    replacing the $b\to u$ tree transition by the $b\to s(d)$ penguin
    transition.
    }}
   \label{fig:2}
\end{figure}

The quark diagrams and the corresponding pole diagrams for decays
of $B$ mesons to the baryonic final state $\B_1\ov\B_2 M$ are more
complicated. In general there are two external $W$-diagrams Figs.
2(a)-2(b), four internal $W$-emissions Figs. 2(c)-2(f), and one
$W$-exchange Fig. 2(g) for the neutral $B$ meson and one
$W$-annihilation Fig. 2(h) for the charged $B$. Because of space
limit, penguin diagrams are not drawn in Fig. 2; they can be
obtained from Figs. 2(c)-2(g) by replacing the $b\to u$ tree
transition  by the $b\to s(d)$ penguin transition. Under the
factorization hypothesis, the relevant factorizable amplitudes are
 \be \label{fact}
 {\rm Figs.~2(a),2(c)}: && \A\propto \la M|(\bar q_3 q_2)|0\ra\la \B_1\ov
 \B_2|(\bar q_1b)|B\ra,  \non \\
 {\rm Figs.~2(b),2(d)}:&& \A\propto \la \B_1\ov \B_2|(\bar q_1 q_2)|0\ra
 \la M|(\bar q_3 b)|B\ra,  \\
 {\rm Figs.~2(g),2(h)}: && \A\propto \la\B_1\ov \B_2 M|(\bar
 q_1q_2)|0\ra\la 0|(\bar q_3 b)|B\ra.  \non
 \en
Since the three-body matrix elements are basically unknown, only
the factorizable amplitudes for Fig. 2(b) or 2(d) are calculable
in practice.

The tree-dominated three-body modes of interest are:
 \be \label{treemodes}
&& \ov B^0\to \pi^+(\rho^+)\{n\bar
 p,~\Lambda\bar \Sigma^-,~\Sigma^0\bar \Sigma^-,~\Sigma^-\bar
 \Lambda,~\Xi^-\bar\Xi^0,~p\bar\Delta^{--},\cdots\}, \non \\
 && \ov B^0\to \pi^-(\rho^-)\{p\bar
 n,~\Sigma^+\bar\Lambda,~\Sigma^+\bar\Sigma^0,~\Lambda\bar\Sigma^+,~\Delta^{++}\bar p,\cdots\},   \\
&&  B^-\to\pi^-(\rho^-)\{ p\bar p,~n\bar
 n,~\Sigma\bar\Sigma,~\Lambda\bar\Lambda,~\Delta\bar\Delta,\cdots\},
 \non
 \en
while some interesting penguin-dominated decays are
 \be \label{penguinmodes}
 && \ov B^0\to \pi^+(\rho^+)\{\Lambda\bar p,~\Sigma^0\bar p,~\Sigma^-\bar
 n,~\Xi^-\bar\Lambda,~\Sigma^+\bar\Delta^{--},\cdots\},  \non \\
 && B^-\to\pi^-(\rho^-)\{ \Sigma^+\bar p,~\Lambda\bar
 n,~\Sigma^0\bar n,~\Xi^0\bar\Lambda,~\Sigma^+\bar\Delta^-,\cdots\},  \non \\
 && \ov B^0\to K^{-(*)}\{ p\bar n,~\Sigma^+\bar\Lambda,~\Sigma^+\bar\Sigma^0,~\Lambda\bar
 \Sigma^+,~\Xi^0\bar\Xi^+,~\Delta^{++}\bar p,\cdots\},  \\
 && B^-\to K^{-(*)}\{p\bar p,~n\bar
 n,~\Sigma\bar\Sigma,~\Lambda\bar\Lambda,~\Delta\bar\Delta,\cdots\},
 \non \\
 && \ov B^0\to \ov K^{0(*)}\{p\bar p,~n\bar
 n,~\Sigma\bar\Sigma,~\Lambda\bar\Lambda,~\Delta\bar\Delta,\cdots\},
 \non \\
 && B^-\to \ov K^{0(*)}\{n\bar
 p,~\Lambda\bar \Sigma^-,~\Sigma^0\bar \Sigma^-,~\Sigma^-\bar
 \Lambda,~p\bar\Delta^{--},\cdots\}. \non
 \en
In the present paper we will focus on octet baryon final states.

To evaluate the factorizable amplitude for Fig. 2(b) or 2(d) we
need to know the octet baryon form factors defined by
 \be
\la \B_1(p_1)\ov\B_2(p_2)|(V\pm A)_\mu|0\ra &=& \bar
u_1(p_1)\Bigg\{f_1^{\B_1\B_2}(q^2)\gamma_\mu+i{f_2^{\B_1\B_2}(q^2)\over
m_1+m_2} \sigma_{\mu\nu}q^\nu+{f_3^{\B_1\B_2}(q^2)\over
m_1+m_2}q_\mu \non  \\ && \pm
\Big[g_1^{\B_1\B_2}(q^2)\gamma_\mu+i{g_2^{\B_1\B_2}(q^2)\over
m_1+m_2} \sigma_{\mu\nu}q^\nu+{g_3^{\B_1\B_2}(q^2)\over
m_1+m_2}q_\mu\Big]\gamma_5\Bigg\}v_2(p_2),
 \en
where $q=p_1+p_2$. For octet baryons one can apply SU(3) symmetry
to relate the vector form factors $f_i^{\B_1\B_2}$ to the nucleon
magnetic and electric form factors. In general, SU(3) symmetry
implies
 \be \label{CG}
 f_i^{\B_1\B_2}(q^2)=d^{\B_1\B_2}D^V_i(q^2)+f^{\B_1\B_2}F^V_i(q^2),
 \qquad g_i^{\B_1\B_2}(q^2)
 =d^{\B_1\B_2}D^A_i(q^2)+f^{\B_1\B_2}F^A_i(q^2),
 \en
where $d^{\B_1\B_2}$ and $f^{\B_1\B_2}$ are the well-known
Clebsch-Gordon coefficients and $F_i^V(q^2)$ and $D_i^V(q^2)$ are
reduced form factors. The nucleon matrix element of the
electromagnetic current is given by
 \be \label{em}
 \la N(p_1)\ov N(p_2)|J_\mu^{\rm em}|0\ra=\bar
 u_N(p_1)\Big[F_1(q^2)\gamma_\mu+i{F_2(q^2)\over 2m_N}\sigma_{\mu\nu}q^\nu
 \Big]v_{\bar N}(p_2).
 \en
Since $J_\mu^{\rm em}=V_\mu^3+{1\over\sqrt{3}}V_\mu^8$, SU(3)
symmetry allows us to determine $F_i^V(q^2)$ and $D_i^V(q^2)$
separately. The results are (see e.g. \cite{Bailin})
 \be \label{vectorF}
 F_{1,2}^V(t)=F_{1,2}^p(t)+{1\over 2}F_{1,2}^n(t), \qquad
 D_{1,2}^V(t)=-{3\over 2}F_{1,2}^n(t), \qquad F_3^V(t)=D_3^V(t)=0,
 \en
with $t\equiv q^2$. At $t=0$ we have
 \be
 F_1^V(0)=1, \qquad D_1^V(0)=0, \qquad F_2^V(0)=\kappa_p+{1\over
 2}\kappa_n, \qquad D_2^V(0)=-{3\over 2}\kappa_n,
 \en
where $\kappa_p=1.79$ and $\kappa_n=-1.91$ are the anomalous
magnetic moments of the proton and neutron, respectively.

The experimental data are customarily described in terms of the
electric and magnetic Sachs form factors $G_E^N(t)$ and $G_M^N(t)$
which are related to $F_1^N$ and $F_2^N$ via
 \be
 G_E^{p,n}(t)=F_1^{p,n}(t)+{t\over 4m_N^2}F_2^{p,n}(t), \qquad G_M^{p,n}(t)
 =F_1^{p,n}(t)+ F_2^{p,n}(t).
 \en
A recent phenomenological fit to the experimental data of nucleon
form factors has been carried out in \cite{CHT1} using the
following parametrization:
 \be \label{GMN}
 |G_M^p(t)| &=& \left({x_1\over t^2}+{x_2\over t^3}+{x_3\over t^4}
 +{x_4\over t^5}+{x_5\over t^6}\right)\left[\ln{t\over
 Q_0^2}\right]^{-\gamma},  \non \\
|G_M^n(t)| &=& \left({y_1\over t^2}+{y_2\over
t^3}\right)\left[\ln{t\over Q_0^2}\right]^{-\gamma},
 \en
where $Q_0=\Lambda_{\rm QCD}\approx$ 300 MeV and $\gamma=2+{4\over
3\beta}=2.148$\,. We will follow \cite{CHT1} to use the best fit
values
 \be
&& x_1=429.88\,{\rm GeV}^4, \qquad x_2=-10783.69\,{\rm GeV}^6,
 \qquad x_3=109738.41\,{\rm GeV}^8, \non \\
&& x_4=-448583.96\,{\rm GeV}^{10}, \qquad x_5=635695.29\,{\rm
GeV}^{12},
 \en
and
 \be \label{GMn1}
 y_1=236.69\,{\rm GeV}^4, \qquad\quad y_2=-579.51\,{\rm GeV}^6,
 \en
extracted from neutron data under the assumption
$|G_E^n|=|G_M^n|$. Note that the form factors given by Eq.
(\ref{GMN}) do satisfy the constraint from perturbative QCD in the
limit of large $t$ \cite{CHT1}. Also as stressed in \cite{CHT1},
time-like magnetic form factors are expected to behave like
space-like magnetic form factors, i.e. real and positive for the
proton, but negative for the neutron.

A new empirical fit to the reanalyzed data for $G_M^p(t)$ in the
region $0<t<30\,{\rm GeV}^2$ is recently given in \cite{Brash}:
 \be \label{BKLH}
 G_M^p(Q^2)={\mu_p\over 1+z_1Q+z_2Q^2+z_3Q^3+z_4Q^4+z_5Q^5},
 \en
with
 \be
 && z_1=(0.116\pm0.040)\,{\rm GeV}^{-1}, \quad z_2=(2.874\pm0.0.098)\,{\rm GeV}^{-2}, \quad
 z_3=(0.241\pm  0.107)\,{\rm GeV}^{-3}, \non \\
 && z_4=(1.006\pm0.069)\,{\rm GeV}^{-4}, \quad z_5=(0.345\pm 0.017)\,{\rm
 GeV}^{-5}\,,
 \en
and $\mu_p=2.79$. An empirical fit to the proton electromagnetic
form factor ratio is also presented in \cite{Brash}
 \be \label{GEp}
 \mu_p{G_E^p(t)\over G_M^p(t)}=\,1.0-(0.130\pm0.005)[t-(0.04\pm
 0.09)],
 \en
for the range $0.04<t<5.6\,{\rm GeV}^2$, indicating that the form
factor ratio decreases with increasing $Q^2$.

As for the axial form factors, no useful information can be
extracted from SU(3) symmetry. Nevertheless, perturbative QCD
indicates that, in the range of high $Q^2$, the form factors
$f_1(t)$ and $g_1(t)$ dominate at $t\to\infty$ and all others are
suppressed by powers of $m/Q$ \cite{Brodsky}. Moreover, all
octet-octet and octet-decuplet form factors at large $t$ can be
related to the magnetic form factors of the nucleon $G_M^p$ and
$G_M^n$ (see Tables II-IV of \cite{Brodsky}). Hence, the axial
form factor $g_1$ at large momentum transfer is fixed.

\subsection{Tree-dominated three-body decays}
 \vskip 0.6 cm
\centerline{ \underline{1.~~$\ov B^0\to n\bar p\pi^+(\rho^+)$}}
 \vskip 0.4cm

This decay receives factorizable contributions from Figs. 2(b),
2(d) with $b\to d$ penguin transition, 2(g) and a nonfactorizable
contribution from Fig. 2(e). As the two-body baryonic decay, we
can neglect the $W$-exchange contribution. Moreover, we shall
assume that this mode is dominated by the factorizable term from
Fig. 2(b) as it is governed by the parameter $a_1$:
 \be
 \A(\ov B^0\to n\bar p\pi^+(\rho^+))_{\rm
 fact}={G_F\over\sqrt{2}}V_{ud}V_{ub}^* a_1\la \pi^+(\rho^+)|(\bar
 ub)_\vma|\ov B^0\ra\la n\bar p|(\bar du)_\vma|0\ra,
 \en
where $a_1=c_1^{\rm eff}+c_2^{\rm eff}/3$ and we have neglected
penguin contributions because the penguin Wilson coefficients
$c_3,\cdots,c_{10}$ are numerically very small. The two-body meson
matrix elements are parametrized in terms of the form factors
$F_0$ and $F_1$ for $B-\pi$ transition
 \be \label{formBpi}
 \la\pi^+(p_\pi)|(\bar ub)_\vma|\ov B^0(p_B)\ra=F_1^{B\pi}
 (q^2)(p_B+p_\pi)_\mu+\left(F_0^{B\pi}
 (q^2)-F_1^{B\pi}(q^2)\right){m_B^2-m_\pi^2\over q^2}q_\mu,
 \en
and form factors $V,A_0,A_1,A_2$ for $B-\rho$ transition
 \be \label{formBrho}
 \la\rho^+(p_\rho)|(\bar ub)_\vma|\ov B^0(p_B)\ra &=& {2\over
 m_B+m_\rho}\epsilon_{\mu\nu\alpha\beta}
\vp^{*\nu}p^\alpha_B p^\beta_\rho
V^{B\rho}(q^2)-i\Bigg\{(m_B+m_\rho) \vp^*_\mu A_1^{B\rho}(q^2) \\
&-& {\vp^*\cdot p_B\over m_B+m_\rho}(p_B+p_\rho)_\mu
A_2^{B\rho}(q^2)-2m_\rho {\vp^*\cdot p_B\over
q^2}q_\mu\left[A_3^{B\rho}(q^2)-A_0^{B\rho}(q^2)\right]\Bigg\},
\non
 \en
with $q=p_B-p_{\pi(\rho)}$ and
 \be \label{A3}
A_3^{B\rho}(q^2)={m_B+m_\rho\over
2m_\rho}A_1^{B\rho}(q^2)-{m_B-m_\rho\over
2m_\rho}A_2^{B\rho}(q^2).
 \en
The factorizable amplitude for the pion emission reads
  \be
 \A(\ov B^0\to n\bar p\pi^+)_{\rm fact}={G_F\over\sqrt{2}}V_{ud}V_{ub}^*\,a_1\bar
 u_{n}\left[(ap\!\!\!/_\pi+b)-(cp\!\!\!/_\pi+d)\gamma_5\right]v_{\bar p},
 \en
where
 \be
 a&=& 2f_1^{np}(t)F_1^{B\pi}(t)+4f_2^{np}(t)F_1^{B\pi}(t), \non \\
 b&=& -2f_2^{np}(t)F_1^{B\pi}(t)(p_n-p_{\bar p})\cdot p_\pi/(2m_N)
 +f_3^{np}(t)F_0^{B\pi}(t)(m_B^2-m_\pi^2)/(2m_N), \non \\
 c &=& 2g_1^{np}(t)F_1^{B\pi}(t), \non \\
 d &=& 2m_Ng_1^{np}(t)\left[F_1^{B\pi}(t)+
 (F^{B\pi}_0(t)-F^{B\pi}_1(t)){m_B^2-m_\pi^2\over
 t}\right]  \non \\
 &&-2g_2^{np}(t)F_1^{B\pi}(t)(p_{n}-p_{\bar p})\cdot
 p_\pi/(2m_N)+g_3^{np}(t)F_0^{B\pi}(t)(m_B^2-m_\pi^2)/(2m_N),
 \en
and $t\equiv q^2=(p_B-p_\pi)^2=(p_n+p_{\bar p})^2$. The amplitude
for the $\rho$ meson case is more cumbersome.

Since the relevant Clebsch-Gordon coefficients are $d^{np}=f^{n
p}=1$, it follows from Eqs. (\ref{CG}) and (\ref{vectorF}) that
the weak form factors have the form
 \be
 f^{np}_{1,2}(t)=F_{1,2}^V(t)+D_{1,2}^V(t)=F_{1,2}^p(t)-F_{1,2}^n(t).
 \en
In terms of the nucleon magnetic and electric form factors, the
weak form factors read
 \be \label{f12np}
 f^{np}_1(t) &=& {{t\over 4m_N^2}G_M^p(t)-G_E^p(t)\over t/(4m_N^2)-1}-
 {{t\over 4m_N^2}G_M^n(t)-G_E^n(t)\over t/(4m_N^2)-1}, \non \\
 f^{np}_2(t) &=& -{G_M^p(t)-G_E^p(t)\over t/(4m_N^2)-1}+
 {G_M^n(t)-G_E^n(t)\over t/(4m_N^2)-1}.
 \en
According to perturbative QCD, the weak form factors in the large
$t$ limit have the expressions \cite{Brodsky}
 \be \label{larget}
 f^{np}_1(t) \to G_M^p(t)-G_M^n(t), \qquad
 g^{np}_1(t) \to {5\over 3}G_M^p(t)+G_M^n(t).
 \en
It is easily seen that this is consistent with the large $t$
behavior of $f_1^{np}$ given by Eq. (\ref{f12np}).

The total decay rate for the process $\ov B^0(p_B)\to n(p_1)+\bar
p(p_2)+\pi^+(p_3)$ is computed by
 \be \Gamma = {1\over (2\pi)^3}\,{1\over
32m_B^3}\int |\A|^2dm_{12}^2dm_{23}^2,
 \en
where $m_{ij}^2=(p_i+p_j)^2$ with $p_3=p_\pi$. To make a numerical
estimate, we apply two different empirical fits of $G_M^p(t)$: Eq.
(\ref{GMN}) denoted by CHT (Chua-Hou-Tsai) and Eq. (\ref{BKLH})
denoted by BKLH (Brash-Kozlov-Li-Huber). For the proton electric
form factor, we shall follow \cite{CHT1} to assume
$|G_E^p(t)|=|G_M^p(t)|$ for CHT form factors and Eq. (\ref{GEp})
for BKLH form factors. That is, we assume that Eq. (\ref{GEp}) is
applicable also to the large $t$ region. As for $B-\pi(\rho)$ form
factors, we consider two distinct models: the Bauer-Stech-Wirbel
(BSW) model \cite{BSW} and the Melikhov-Stech (MS) model based on
the constituent quark picture \cite{Melikhov}.\footnote{The QCD
sum rule method  based on the light-cone sum rule analysis
\cite{LCSR} is also one of the popular form-factor models.
However, we found that some divergence occurs in the phase space
integration when applying this model.} The BSW model assumes a
monopole behavior for all the form factors. However, this is not
consistent with heavy quark symmetry for heavy-to-light
transitions. For example, the form factors $F_1,V,A_0,A_2$ in the
infinite quark mass limit should have the same $q^2$ dependence
and they differ from $F_0$ and $A_1$ by an additional pole factor
\cite{Xu}. Nevertheless, we apply this model for comparison.

Considering only the vector-current contribution to baryon matrix
element, we obtain the results shown in the first entry of Table
III. Our calculations are in agreement with \cite{CHT2} when the
BSW model and CHT form factors are used. However, we see from
Table III that the branching ratio for $\ov B^0\to n\bar p\rho^+$
in the BSW model is slightly larger. This is ascribed to the
monopole form factor $q^2$ dependence for all the $B-\rho$ form
factors. If one changes the form factor momentum dependence from
monopole to dipole form for $A_1$ and $V$ (sometimes referred to
as the BSWII model in the literature), the resulting branching
ratios are very similar to that in the MS model.

\begin{table}[ht]
\caption{Branching ratios of $\ov B^0\to n\bar p\pi^+(\rho^+)$ in
two different form-factor models for $B-\pi(\rho)$ transition. Two
distinct empirical fits for the proton magnetic form factor given
in Eqs. (\ref{GMN}) and (\ref{BKLH}), denoted by CHT and BKLH
respectively, are utilized. The neutron form factors are taken
from (\ref{GMN}) and (\ref{GMn1}). Branching ratios in the first
entry are without contributions from the axial form factors
$g^{np}_i(t)$ and those in the second entry take into account
contributions  from the asymptotic form factor $g^{np}_1(t)$ given
by Eq. (\ref{larget}).}
\begin{center}
\begin{tabular}{l| c c c c  }
& \multicolumn{2}{c}{$G_M^p$ (CHT)} & \multicolumn{2}{c}{$G_M^p$
(BKLH)}
\\ \cline{2-3} \cline{4-5} \raisebox{1.5ex}[0cm][0cm]{}
 & MS  & BSW & MS  & BSW   \\ \hline
 $\ov B^0\to n\bar p\pi^+$ & $1.7\times 10^{-6}$  & $1.8\times
 10^{-6}$ & $8.0\times 10^{-7}$  & $8.5\times 10^{-7}$ \\
 & $1.7\times 10^{-6}$  & $1.9\times
 10^{-6}$ & $1.1\times 10^{-6}$  & $1.3\times 10^{-6}$ \\
 $\ov B^0\to n\bar p\rho^+$ & $3.3\times 10^{-6}$  & $4.8\times
 10^{-6}$ & $4.2\times 10^{-6}$  & $5.5\times 10^{-6}$ \\
 & $3.4\times 10^{-6}$  & $4.9\times
 10^{-6}$ & $4.6\times 10^{-6}$  & $5.9\times 10^{-6}$ \\
\end{tabular}
\end{center}
\end{table}

To estimate the contribution from the axial vector current, we
might assume that $g_1(t)$ takes the asymptotic form ${5\over
3}G_M^p(t)+G_M^n(t)$ [see Eq. (\ref{larget})]. It turns out that
the contribution due to $g_1(t)$ is very small for the CHT form
factor $G_M^p$ but not negligible for $G_M^p$(BKLH). It is
interesting to notice that the rate of $n\bar p\rho^+$ is larger
than that of $n\bar p\pi^+$ by a factor of $2\sim 3$ if the CHT
parametrization for $G_M^p$ is employed, whereas the ratio becomes
as large as 5 for $G_M^p$(BKLH).

Since both $B^0$ and $\ov B^0$ can decay into $n\bar
p\pi^+(\rho^+)$, experimentally one has to disentangle the
``background" contribution from the $B^0-\bar B^0$ mixing or to
tag the $B$ meson. Therefore, we will give an estimate of $\ov
B^0\to p\bar n\pi^-(\rho^-)$ next.

\vskip 0.6 cm \centerline{ \underline{2.~~$\ov B^0\to p\bar
n\pi^-(\rho^-)$}}
 \vskip 0.4cm

This decay receives contributions from Figs. 2(a), 2(e) and 2(g).
As the previous decay, we will assume that it is dominated by the
factorizable contribution from Fig. 2(a). Unfortunately,  as shown
in Eq. (\ref{fact}), it involves a three-body matrix element that
cannot be evaluated directly. Instead, we will evaluate the
low-lying pole diagrams with the strong process $\ov B^0
\to\{\Lambda_b^{(*)},\Sigma_b^{0(*)}\}\bar n$ followed by the weak
decays $\{\Lambda_b^{(*)},\Sigma_b^{0(*)}\}\to
\pi^-p$.\footnote{There is another pole diagram with the weak
decay $\ov B^0\to \pi^-\pi^+(\rho^+)$ followed by the strong
process $\pi^+(\rho^+)\to p\bar n$ [see Fig. 2(a)]. However, this
pole amplitude is expected to be suppressed as the intermediate
pion state is far off its mass shell. Consequently, the
nucleon-nucleon-pion coupling is subject to a large suppression
due to the form-factor effects at large $q^2$.} Consider the
${1\over 2}^+$ intermediate poles. Applying factorization to
$\Lambda_b\to \pi^-p$ yields the pole amplitude
 \be
 \A(\ov B^0\to p\bar n\pi^-) &=& -{G_F\over\sqrt{2}}V_{ud}V_{ub}^*\,
 g_{\Lambda_b\to \ov B^0n}f_\pi\,a_1\,\bar
 u_{p}\Bigg\{f_1^{\Lambda_b p}(m_\pi^2)\Big[2p_\pi\cdot
 p_{p}+p\!\!\!/_\pi(m_{\Lambda_b}-m_{p})\Big]\gamma_5  \non \\
 && +g_1^{\Lambda_b p}(m_\pi^2)\Big[2p_\pi\cdot
 p_{p}-p\!\!\!/_\pi(m_{\Lambda_b}+m_{p})\Big]\Bigg\}
 v_{\bar n}\times{1\over
 (p_{p}+p_\pi)^2-m_{\Lambda_b}^2 } \non \\ && +~~(\Lambda_b\to \Sigma_b^0),
 \en
where we have employed the heavy-light baryon form factors defined
by
 \be
 \la p(p_{p})|(\bar ub)_{_{V\pm A}}|\Lambda_b(p_{\Lambda_b})\ra &=&
\bar u_p\Bigg\{f_1^{\Lambda_b p}(p_\pi^2)\gamma_\mu
+i{f_2^{\Lambda_b p}(p_\pi^2)\over m_{\Lambda_b}+m_p }
 \sigma_{\mu\nu}p_\pi^\nu+{f_3^{\Lambda_b p}(p_\pi^2)\over m_{\Lambda_b}+m_p}
 p_{\pi\mu} \non  \\ &&
 \pm\Big[g_1^{\Lambda_b p}(p_\pi^2)\gamma_\mu+i{g_2^{\Lambda_b p}(p_\pi^2)\over
 m_{\Lambda_b}+m_p} \sigma_{\mu\nu}p_\pi^\nu+{g_3^{\Lambda_b p}(p_\pi^2)\over
 m_{\Lambda_b}+m_p}p_{\pi\mu}\Big] \gamma_5\Bigg\}u_{\Lambda_b},
 \en
with $p_\pi=p_{\Lambda_b}-p_{p}$.

For the heavy-light form factors $f_i^{\B_1\B_2}$ and
$g_i^{\B_1\B_2}$, we will follow \cite{CT96} to apply the
nonrelativistic quark model to evaluate the weak current-induced
baryon form factors at zero recoil in the rest frame of the heavy
parent baryon, where the quark model is most trustworthy. This
quark model approach has the merit that it is applicable to
heavy-to-heavy and heavy-to-light baryonic transitions at maximum
$q^2$.  Following \cite{Cheng97} we have\footnote{The form factors
for $\Lambda_b-p$ transition at $q^2=0$ are given in Table I of
\cite{Cheng97}. For $\Sigma_b^0-p$ form factors at zero recoil, it
can be evaluated using Eq. (22) of \cite{CT96}. Note that the spin
factor is $\eta=-{1\over 3}$ and the flavor factor is
$N_{\Sigma_b^0p}=1/\sqrt{6}$ for $\Sigma_b^0-p$ transition.}
 \be \label{Lambdabp}
&& f_1^{\Lambda_b p}(q^2_m)=g_1^{\Lambda_bp}(q^2_m)=0.86,
 \quad
 f_2^{\Lambda_b p}(q^2_m)=g_3^{\Lambda_bp}(q^2_m)=-0.51,
 \non \\
&& f_3^{\Lambda_b p}(q^2_m)=g_2^{\Lambda_b p}(q^2_m)=-0.22,
 \en
for $\Lambda_b-p$ transition at zero recoil
$q_m^2=(m_{\Lambda_b}-m_p)^2$, and
 \be \label{Sigmabp}
&& f_1^{\Sigma_b^0 p}(q^2_m)=\,1.65,  \qquad
 f_2^{\Sigma_b^0 p}(q^2_m)=\,1.92, \qquad f_3^{\Sigma_b^0
 p}(q^2_m)=-1.72,  \non \\
 && g_1^{\Sigma_b^0 p}(q^2_m)=-0.17,  \qquad
 g_2^{\Sigma_b^0 p}(q^2_m)=\,0.04, \qquad g_3^{\Sigma_b^0
 p}(q^2_m)=0.10,
 \en
for $\Sigma_b^0-p$ transition at $q^2_m=(m_{\Sigma_b}-m_p)^2$.
Since the calculation for the $q^2$ dependence of form factors is
beyond the scope of the non-relativistic quark model, we will
follow the conventional practice to assume a pole dominance for
the form-factor $q^2$ behavior:
 \be
 f(q^2)=f(q^2_m)\left({1-q^2_m/m^2_V\over 1-q^2/m_V^2} \right)^n\,,\qquad
 g(q^2)=g(q^2_m)
\left({1-q^2_m/m^2_A\over 1-q^2/m_A^2} \right)^n\,,
 \en
where $m_V$ ($m_A$) is the pole mass of the vector (axial-vector)
meson with the same quantum number as the current under
consideration. The function
 \be
 G(q^2)=\left({1-q^2_m/m^2_{\rm
pole}\over 1-q^2/m_{\rm pole}^2} \right)^n
 \en
plays the role of the baryon Isgur-Wise function $\zeta(\omega)$
for $\Lambda_Q\to \Lambda_{Q'}$ transition, namely, $G=1$ at
$q^2=q^2_m$. Previous model calculations of $\zeta(\omega)$
\cite{Jenkins,Sadzi,GuoK,Dai,Iva97} indicates that it is
consistent with $G(q^2)$ with $n=2$. However, a recent calculation
of $\zeta(\omega)$ in \cite{Iva99} yields
 \be
 \zeta(\omega)=\left({2\over 1+\omega}\right)^{1.23+0.4/\omega}
 \en
and this clearly favors $n=1$. As we shall below, the recent first
observation of $B^-\to p\bar p K^-$ by Belle \cite{Bellebaryon}
also favors a monopole $q^2$ dependence for baryon form factors.

The calculation of $\ov B^0\to p\bar n \rho^-$ is similar to that
of $p\bar n \pi^-$ except that the vacuum-$\rho$ matrix element
now reads
 \be
 \la \rho^{-}|\bar d\gamma_\mu u|0\ra &=& f_{\rho}
 m_{\rho}\vp_\mu^*,
 \en
and that the computation is much more tedious than the pion case,
though it is straightforward. Using the pole masses $m_V=5.32$
GeV, $m_A=5.71$ GeV and the decay constant $f_\rho=216$ MeV, we
obtain
 \be
 \B(\ov B^0\to p\bar n\pi^-) &=& 2.8\times 10^{-6}~~(2.7\times
 10^{-7}),  \non \\
 \B(\ov B^0\to p\bar n\rho^-) &=& 8.2\times 10^{-6}~~(8.2\times
 10^{-7}),
 \en
for a monopole (dipole) $q^2$ dependence for baryon form factors.
Since  $g_{\Lambda_b\to\ov B^0n}=-3\sqrt{3}\,g_{\Sigma_b^0\to\ov
B^0n}$ [cf. Eq. (\ref{grel3})], the contribution due to the
$\Lambda_b$ and $\Sigma_b^0$ poles is destructive. In the
calculation we have used $|g_{\Lambda_b\to\ov B^0n}|=16$
\cite{CYBbaryon}.

Three remarks are in order. First, in the calculation we have
neglected other nonfactorizable contributions from Fig. 2(e). For
the pole diagrams, we did not evaluate the ${1\over 2}^-$ pole
contributions owing to the technical difficulties for the bag
model in dealing with the negative-parity baryon states. Second,
since $n=1$ is favored by the recent measurement of the decay
$B^-\to p\bar p K^-$, as we shall see below, it turns out that
$B^0\to n\bar p\rho^+$ has a large branching ratio of order
$1\times 10^{-5}$ for $n=1$. Third, the decay $\ov B^0\to p\bar
n\pi^-$ receives the resonant contribution $\ov B^0\to
p\bar\Delta^-$ followed by the strong decay $\bar\Delta^-\to \bar
n\pi^-$. Since the branching ratio for $\ov B^0\to p\bar\Delta^-$
is only of order $6\times 10^{-8}$ (see Table II), the resonant
contribution due to the $\Delta$ is thus negligible.

 \vskip 0.6 cm
\centerline{ \underline{3.~~$\ov B^0\to
\{\Sigma^0\bar\Sigma^-,~\Sigma^-\bar\Lambda,~\Xi^-\bar\Xi^0\}\pi^+(\rho^+)$}}
 \vskip 0.4cm

The calculation  for the decays $\ov B^0\to
\{\Sigma^0\bar\Sigma^-,~\Sigma^-\bar\Lambda,~\Xi^-\bar\Xi^0\}\pi^+(\rho^+)$
is the same as that for $\ov B^0\to n\bar p\pi^+(\rho^+)$ except
for different baryonic form factors in the final states. The
relevant Clebsch-Gordon coefficients for weak form factors are
(see e.g. \cite{Bailin})
 \be
 && d^{\Lambda\Sigma^+}=\,\sqrt{2\over 3}, \qquad
 d^{\Sigma^0\Sigma^+}=0, \qquad
 d^{\Xi^0\Xi^+}=1, \non \\
 && f^{\Lambda\Sigma^+}=0,\qquad f^{\Sigma^0\Sigma^+}=\sqrt{2},
 \qquad f^{\Xi^0\Xi^+}=-1.
 \en
Then (\ref{CG}) and (\ref{vectorF}) lead to
 \be
&& f_{1,2}^{\Lambda\Sigma^+}(t)=-\sqrt{3\over 2}\,F_{1,2}^n(t),
 \qquad  \qquad f_{1,2}^{\Xi^0\Xi^+}=-F_{1,2}^p(t)-2F_{1,2}^n(t),
 \non \\ &&
 f_{1,2}^{\Sigma^0\Sigma^+}(t)=\sqrt{2}F_{1,2}^p(t)+{1\over\sqrt{2}}F_{1,2}^n(t).
 \en
A straightforward calculation gives $\B(\ov
B^0\to\Sigma^-\bar\Lambda\pi^+)=2.9\times 10^{-7}$, $\B(\ov B^0\to
\Xi^-\bar\Xi^0\pi^+)=2.0\times 10^{-7}$ and $\B(\ov B^0\to
\Sigma^0\bar\Sigma^-\pi^+)=6.4\times 10^{-9}$. Compared to the
$n\bar p\pi^+$ mode, the decay rates of above three decays are
suppressed owing to smaller baryon form factors and less
three-body phase spaces available.

 \vskip 0.6 cm
\centerline{ \underline{4.~~$B^-\to\{p\bar p,n\bar
n,\Sigma^+\bar\Sigma^-,\cdots\}\pi^-(\rho^-)$}}
 \vskip 0.4cm

Let us first consider the decay $B^-\to p\bar p\pi^-$. It receives
factorizable contributions from Figs. 2(a) and 2(d):
 \be \label{factamp}
 A(B^-\to p\bar p\pi^-)_{\rm fact} &=&
 {G_F\over\sqrt{2}}V_{ub}V_{ud}^*\Big\{a_1\la \pi^-|(\bar
 du)_\vma|0\ra\la p\bar p|(\bar ub)_\vma|B^-\ra \non \\
 &+& a_2\la\pi^-|(\bar db)_\vma|B^-\ra\la p\bar p|(\bar
 uu)_\vma|0\ra\Big\} \equiv  A_1+A_2,
 \en
where $a_{1,2}=c_{1,2}^{\rm eff}+c_{2,1}^{\rm eff}/3$. In analog
to the previous mode, we will evaluate the corresponding low-lying
pole diagrams for the factorizable external $W$-emission
amplitude, namely, the strong process $B^-\to
\{\Lambda_b^{(*)},\Sigma_b^{0(*)}\}\bar p$, followed by the weak
decay $\{\Lambda_b^{(*)},\Sigma_b^{0(*)}\}\to p\pi^-$. Its
amplitude governed by the ${1\over 2}^+$ poles is given by
 \be
 A_1 &=& -{G_F\over\sqrt{2}}V_{ub}V_{ud}^*\,g_{\Lambda_b\to B^-p}f_\pi\,a_1\,\bar
 u_{p}\Big\{f_1^{\Lambda_b p}(m_\pi^2)[2p_\pi\cdot
 p_{p}+p\!\!\!/_\pi(m_{\Lambda_b}-m_{p})]\gamma_5  \non \\
 &+& g_1^{\Lambda_b p}(m_\pi^2)[2p_\pi\cdot
 p_{p}-p\!\!\!/_\pi(m_{\Lambda_b}+m_{p})]\Big\}
 v_{\bar p}\times{1\over
 (p_{p}+p_\pi)^2-m_{\Lambda_b}^2 } \non \\
 &+& ~~(\Lambda_b\to\Sigma_b^0),
 \en
where we have applied factorization to the weak decay
$\{\Lambda_b,\Sigma_b^0\}\to p\pi^-$. To evaluate the factorizable
amplitude $A_2$, we apply the isospin symmetry
relations\footnote{This isospin relation amounts to assuming $\la
N|(\bar ss)_\vma|N\ra=0$, an assumption supported by the OZI
rule.}
 \be
 \la n|(\bar uu)_\vma|n\ra=\la p|(\bar dd)_\vma|p\ra, \qquad
 \la n|(\bar dd)_\vma|n\ra=\la p|(\bar uu)_\vma|p\ra,
 \en
to relate the form factors $f_1^{pp}$ and $f_2^{pp}$ appearing in
the vector current $p\bar p$ matrix element
 \be
  \la p(p_1)\bar p(p_2)|\bar u\gamma_\mu u|0\ra =\bar
  u_p(p_1)\Big[f_1^{pp}(q^2)\gamma_\mu+i{f_2^{pp}(q^2)\over
  2m_p}\sigma_{\mu\nu}q^\nu\Big]v_{\bar p}(p_2),
  \en
to the electromagnetic form factors $F_1$ and $F_2$ defined in the
nucleon matrix element Eq. (\ref{em}). We find
 \be
 f_1^{pp}(t)=2F_1^p(t)+F_1^n(t), \qquad
 f_2^{pp}(t)=2F_2^p(t)+F_2^n(t).
 \en

A straightforward calculation indicates that the contribution from
$a_2$ is small and negligible due mainly to the small vector form
factors $f_{1,2}^{pp}$. The $a_1$ contribution gives a branching
ratio of order $3.8\times 10^{-6}$ for $n=1$ and $2.7\times
10^{-7}$ for $n=2$. As we shall see below, as far as the
factorizable $a_1$ contribution is concerned, the tree-dominated
$B^-\to p\bar p\pi^-$ and the penguin-dominated decay $B^-\to
p\bar p K^-$ have almost the same rate and the latter has been
observed recently \cite{Bellebaryon}. In some sense this is very
similar to the mesonic decays $B\to K\pi$ and $\pi\pi$. Without
the chiral enhancement for penguin contributions, one will have
$\pi\pi> K\pi$. The experimental observation \cite{Kpi} that
$K^-\pi^+>\pi^-\pi^+$ and $\ov K^0\pi^->\pi^0\pi^-$ clearly
implies the importance of penguin chiral enhancement. It is quite
possible that for baryonic $B$ decay we also have $p\bar p
K^->p\bar p\pi^-$. Note that the $a_2$ contribution to $p\bar
p\pi^-$ is destructive and it is subject to many uncertainties.
For example, the axial-vector current contribution to the $p\bar
p$ matrix element has been neglected so far and the value of $a_2$
is numerically very small if $a_2=c_2^{\rm eff}+c_1^{\rm eff}/3$.
A large value of $a_2$ of order $0.40-0.55$ \cite{Cheng01}, as
indicated by the recent observation of $\ov B^0\to D^0\pi^0$
\cite{BelleCLEO}, and an inclusion of axial form factor
contributions may suppress $p\bar p\pi^-$ relative to $p\bar
pK^-$. Another effect we have neglected thus far is the penguin
contribution. Just as the $B\to\pi\pi$ decay, the tree-penguin
interference for $B^-\to p\bar p\pi^-$ may turn out to be
destructive for a certain range of the phase angle $\gamma$. In
view of the aforementioned considerations, we will prefer to carry
out a full analysis of $B^-\to\{p\bar p,n\bar
n,\Sigma^+\bar\Sigma^-,\cdots\}\pi^-(\rho^-)$ decays in a separate
publication. It appears to us that $B^-\to p\bar p\pi^-$ should
have a branching ratio at least of order $10^{-6}$, based on the
recent measurement of $B^-\to p\bar pK^-$ to be discussed below.

Thus far we have focused on the nonresonant decay of $p\bar
p\pi^-$. It also receives resonant contributions, for example
$B^-\to p\bar\Delta^{--}$ and $B^-\to \bar p N^0(1440)$. As
discussed in Sec. III.A, the branching ratio of the former is of
order $(1-2)\times 10^{-6}$, to be compared with the recent
measurement by Belle \cite{Bellebaryon}
 \be
 \B(B^-\to p\bar p\pi^-)=(1.9^{+1.0}_{-0.9}\pm0.3)\times 10^{-6}.
 \en
Therefore, the direct nonresonant contribution is probably smaller
than the resonant ones. Experimentally, it is thus important to
study the resonance effects through the Dalitz plot analysis.

\subsection{Penguin-dominated three-body decays}

\vskip 0.6cm \centerline{ \underline{1.~~$\ov B\to N\ov N
K^{(*)}$}} \vskip 0.4cm

The decay $B^-\to p\bar p K^{-(*)}$  is mainly governed by the
diagrams Figs. 2(a) and 2(c) with the factorizable amplitude
 \be \label{ppK}
 \A(B^-\to p\bar p K^{-(*)})_{\rm fact} &=&  {G_F\over\sqrt{2}}\Bigg\{V_{ub}V_{us}^*\Big[ a_1
 \la K^{-(*)}|(\bar su)_\vma|0\ra\la p\bar p|(\bar  ub)_\vma|B^-\ra \non \\
 &+& a_2\la p\bar p|(\bar uu)_\vma|0\ra\la K^{-(*)}|(\bar
 sb)_\vma|B^-\ra \non \\
 &-& V_{tb}V^*_{ts}\Big[ a_3\la p\bar p|(\bar uu+\bar
 dd+\bar ss)_\vma|0\ra\la K^{-(*)}|(\bar sb)_\vma|B^-\ra \non \\
 &+& a_5\la p\bar p|(\bar uu+\bar
 dd+\bar ss)_\vpa|0\ra\la K^{-(*)}|(\bar sb)_\vma|B^-\ra \non \\
 &+& (a_4+a_{10})\la K^{-(*)}|(\bar su)_\vma|0\ra\la p\bar p|(\bar
 ub)_\vma|B^-\ra  \non \\  &-& 2(a_6+a_8)\la K^{-(*)}|
 \bar s(1+\gamma_5)u|\ov 0\ra\la p\bar p|\bar u(1-\gamma_5)b|B^-\ra \\
&+& (a_4+a_{10})\la K^{-(*)}p\bar p|(\bar
su)_\vma|0\ra\la 0|(\bar  ub)_\vma|B^-\ra  \non \\
  &-&  2(a_6+a_8)\la K^{-(*)}p\bar
 p|\bar s(1+\gamma_5)u|0\ra\la 0|\bar u(1-\gamma_5)b|B^-\ra \non
 \\ &+& a_9\la p\bar p|(\bar uu-{1\over 3}\bar
 dd-{1\over 3}\bar ss)_\vpa|0\ra\la K^{-(*)}|(\bar sb)_\vma|B^-\ra
 \Big]\Bigg\},  \non
 \en
with
 \be
 a_{2i}=c_{2i}^{\rm eff}+{1\over N_c}c_{2i-1}^{\rm eff},\qquad
 a_{2i-1}=c_{2i-1}^{\rm eff}+{1\over N_c}c_{2i}^{\rm
 eff}.
 \en
In Eq. (\ref{ppK}) the last two terms correspond to weak
annihilation. As in the decay $\ov B^0\to p\bar n\pi^-$, since we
do not know how to evaluate the 3-body hadronic matrix element, we
will instead evaluate the corresponding low-lying pole diagrams
with the strong process $B^-
\to\{\Lambda_b^{(*)},\Sigma_b^{0(*)}\}\bar p$ followed by the weak
decays $\{\Lambda_b^{(*)},\Sigma_b^{0(*)}\}\to K^{-(*)}p$ [cf.
Figs. 2(a) and 2(c)]. Consider the ${1\over 2}^+$ intermediate
poles and the final state $K^-$ first. Applying factorization to
$\Lambda_b\to K^-p$ yields
 \be
 \la K^-p|{\cal H}_W|\Lambda_b\ra &=& {G_F\over\sqrt{2}}\Bigg\{\Big[
 V_{ub}V_{us}^*a_1-V_{tb}V_{ts}^*(a_4+a_{10})\Big]\la K^-|(\bar
 su)_\vma|0\ra\la p|(\bar ub)_\vma|\Lambda_b\ra \non \\
 &+& 2V_{tb}V_{ts}^*(a_6+a_8)\,{m_K^2\over m_b m_s}\la K^-|(\bar
 su)_\vpa|0\ra\la p|(\bar ub)_\vpa|\Lambda_b\ra\Bigg\},
 \en
where we have applied equations of motion
 \be
 -i\partial^\mu(\bar q_1\gamma_\mu q_2)=(m_1-m_2)\bar q_1 q_2,
 \qquad\quad -i\partial^\mu(\bar q_1\gamma_\mu\gamma_5 q_2)=(m_1+m_2)\bar q_1\gamma_5
 q_2.
 \en

The pole amplitude then has the form
 \be
 \A(B^-\to p\bar pK^-) &=& -{G_F\over\sqrt{2}}\,g_{\Lambda_b^0\to B^-p}\,f_K\,\bar
 u_{p}\Bigg\{f_1^{\Lambda_b p}(m_K^2)h\Big[2p_K\cdot
 p_{p}+p\!\!\!/_K(m_{\Lambda_b}-m_{p})\Big]\gamma_5  \non \\
 && +g_1^{\Lambda_b p}(m_K^2)h'\Big[2p_K\cdot
 p_{p}-p\!\!\!/_K(m_{\Lambda_b}+m_{p})\Big]\Bigg\}
 v_{\bar p}\times{1\over  (p_{p}+p_ K)^2-m_{\Lambda_b}^2 } \non
 \\  && +~~(\Lambda_b\to \Sigma_b^0),
 \en
with
 \be \label{h}
 h &=& V_{ub}V_{us}^*a_1-V_{tb}V_{ts}^*\Big\{a_4+a_{10}+2(a_6+a_8){m_K^2\over m_b
 m_s}\Big\},  \non \\
 h'&=& V_{ub}V_{us}^*a_1-V_{tb}V_{ts}^*\Big\{a_4+a_{10}-2(a_6+a_8){m_K^2\over m_b
 m_s}\Big\}.
 \en
Since $g_{\Lambda_b\to B^-p}=3\sqrt{3}\,g_{\Sigma_b^0\to B^-p}$
[Eq. (\ref{grel1})], it is evident that the pole contributions
arising from the $\Lambda_b$ and $\Sigma_b^0$ intermediate states
are constructive and dominated by the former one.

The amplitude of $B^-\to p\bar p K^{*-}$ is similar to that of
$p\bar pK^-$ except that there are no $a_6$ and $a_8$ penguin
contributions to $h$ or $h'$ given in Eq. (\ref{h}) owing to the
fact that $\la K^{*-}|\bar su|0\ra=0$. For numerical calculations
of decay rates we use the running quark masses $m_b(m_b)=4.4$ GeV,
$m_s(m_b)=90$ MeV and the decay constant $f_{K^*}=221$ MeV. Note
that the corresponding running strange quark mass at $\mu=1$ GeV
is 140 MeV.  Applying the baryon form factors given by Eqs.
(\ref{Lambdabp}) and (\ref{Sigmabp}) we obtain
 \be
 \B(B^-\to p\bar p K^-) &=& 4.0\times 10^{-6}~~(2.3\times
 10^{-7}),  \non \\
 \B(B^-\to p\bar p K^{*-}) &=& 2.3\times 10^{-6}~~(2.1\times
 10^{-7}),
 \en
for $n=1$ ($n=2$), where use of the strong coupling
$|g_{\Lambda_b^0\to B^-p}|=16$ has been made. As stressed before,
the large chiral enhancement of penguin contributions
characterized by the $m_K^2/(m_bm_s)$ term accounts for the
sizable decay rate of $B^-\to p\bar p K^-$.

An observation of this mode has recently been reported by Belle
\cite{Bellebaryon}
 \be
 \B(B^-\to p\bar p K^-) &=& (4.3^{+1.1}_{-0.9}\pm 0.5)\times
 10^{-6}.
 \en
This is the first ever measurement of the penguin-dominated
charmless baryonic $B$ decay. Evidently, the model prediction is
in good agreement with experiment provided that the baryon form
factor $q^2$ dependence is of the monopole form (i.e. $n=1$).
However, in view of many assumptions and uncertainties involved in
the calculation, the statement about the monopole $q^2$ dependence
for heavy-to-light baryonic form factors should be regarded as a
suggestion rather than a firm one. The absence of penguin
contributions of $a_6$ and $a_8$ to $K^*$ production explains why
the $p\bar pK^{*-}$ rate is smaller than $p\bar p K^-$, contrary
to the case of $\ov B^0\to n\bar p\pi^+(\rho^+)$ where the ratio
of $\rho^+/\pi^+$ can be as large as 5.

In Fig. 3 we show the differential decay rate $d\Gamma/dt$ of
$B^-\to p\bar p K^-$ where $t=(p_p+p_{\bar p})^2=(p_B-p_K)^2$.
Evidently, the spectrum peaks at $t\sim 5.5\,{\rm GeV}^2$,
indicating a threshold enhancement for baryon production and a
fast recoil kaon accompanied by a baryon pair with low invariant
mass.

\begin{figure}[ptb]
\hspace{2cm} \psfig{figure=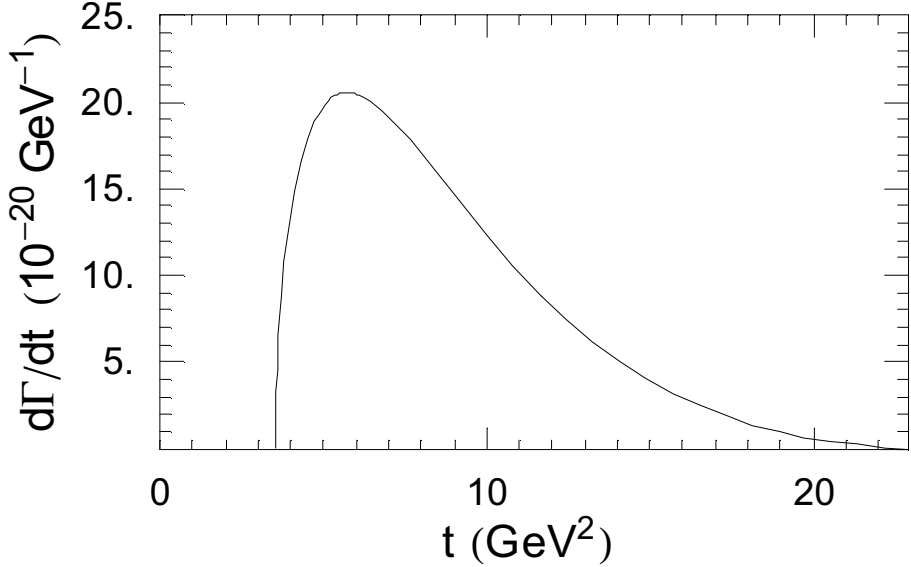,height=3in} \vspace{0.6cm}
    \caption{{\small The differential decay rate of $B^-\to p\bar
    p K^-$ where $t=(p_p+p_{\bar p})^2=(p_B-p_K)^2$.
    }}
\end{figure}

For other $N\ov N K^{(*)}$ modes, it is easily seen that the pole
amplitude of $\ov B^0\to p\bar n K^{-(*)}$ is very similar to that
of $B^-\to p\bar p K^{-(*)}$ except the $\Lambda_b$ and
$\Sigma_b^0$ poles contribute destructively owing to the relation
$g_{\Lambda_b\to \ov B^0n}=-3\sqrt{3}g_{\Sigma_b^0\to \ov B^0n}$
[Eq. (\ref{grel3})]. Repeating the same calculation as before
gives
 \be
 \B(\ov B^0\to p\bar n K^-) &=& 1.9\times 10^{-6}~~(1.5\times
 10^{-7}),  \non \\
 \B(\ov B^0\to p\bar n K^{*-}) &=& 1.8\times 10^{-6}~~(1.9\times
 10^{-7}),
 \en
for $n=1$ ($n=2$). As for $\ov B^0\to n\ov n \bar K^{0(*)}$, its
pole amplitude is the same as $\ov B^0\to p\bar n K^{-(*)}$ except
that the electroweak parameters $a_8$ and $a_{10}$ in Eq.
(\ref{h}) are replaced by $-{1\over 2}a_8$ and $-{1\over
2}a_{10}$, respectively. Since these parameters are very small,
the mode $n\bar n \bar K^{0(*)}$ has a similar rate as $p\bar n
K^{-(*)}$.

As for the decays $\ov B^0\to p\bar p \bar K^{0(*)}$ and $B^-\to
n\bar n K^{-(*)}$, their branching ratios are suppressed, of order
a few times of $10^{-7}$ for $K$ production and $5\times 10^{-8}$
for $K^*$. This is attributed to the fact that only the $\Sigma_b$
pole contributes and its coupling with the $B$ meson and the
nucleon is smaller compared to $\Lambda_b$. The current limit is
$\B(\ov B^0\to p\bar p\ov K^0)<7.2\times 10^{-6}$
\cite{Bellebaryon}.

\vskip 0.6cm \centerline{ \underline{2.~~$\ov B^0\to\Lambda\bar
{\lowercase{p}}\pi^+(\rho^+)$}} \vskip 0.4cm

This decay receives contributions from Figs. 2(b), 2(d), 2(e) and
2(g). The factorizable amplitude from Figs. (2b) and 2(d)
including tree and penguin transitions is
 \be \label{Lambdappi}
 \A(\ov B^0\to\Lambda\bar p\pi^+(\rho^+))_{\rm fact} &=&
 {G_F\over\sqrt{2}}\Bigg\{V_{ub}V_{us}^* a_1\la
 \pi^+(\rho^+)|(\bar ub)_\vma|\ov B^0\ra\la\Lambda\bar p|(\bar
 su)_\vma|0\ra \non \\
 &-& V_{tb}V^*_{ts}\Big[
 (a_4+a_{10})\la\pi^+(\rho^+)|(\bar ub)_\vma|
 \ov B^0\ra\la\Lambda\bar p|(\bar su)_\vma|0\ra  \\
 &-& 2(a_6+a_8)\la\pi^+(\rho^+)|
 \bar u(1-\gamma_5)b|\ov B^0\ra\la\Lambda\bar p|\bar s(1+\gamma_5)u|0\ra
 \non\\  &+& (a_4-{1\over 2}a_{10})\la\pi^+(\rho^+)\Lambda\bar
 p|(\bar sd)_\vma|0\ra\la 0|(\bar db)_\vma|\ov B^0\ra  \non \\
 &-&  2(a_6-{1\over 2}a_8)\la\pi^+(\rho^+)\Lambda\bar
 p|\bar s(1+\gamma_5)d|0\ra\la 0|\bar d(1-\gamma_5)b|\ov B^0\ra
 \Big]\Bigg\},  \non
 \en
where the first term corresponds to external $W$-emission, second
and third terms to the $b\to s$ penguin transition and  the last
two terms to penguin-induced weak annihilation. We shall neglect
the weak-annihilation contributions in the practical calculation.
Applying equations of motion we obtain
 \be
 \la \Lambda\bar p|\bar s(1+\gamma_5)u|0\ra  &=& {(p_\Lambda+p_{\bar
 p})^\mu\over m_s-m_\mu}\la\Lambda\bar p|\bar s\gamma_\mu b|0\ra+
 {(p_\Lambda+p_{\bar
 p})^\mu\over m_s+m_\mu}\la\Lambda\bar p|\bar s\gamma_\mu\gamma_5 b|0\ra
  \non \\
 &=& {m_\Lambda-m_p\over m_s-m_u}f_1^{\Lambda p}(t)\bar u_\Lambda
 v_{\bar p} +{1\over m_s+m_u}\Big[\,(m_\Lambda+m_p)g_1^{\Lambda p}(t) \non \\
 &+& {t\over
 m_\Lambda+m_p}g_3^{\Lambda p}(t)\Big]\bar u_\Lambda\gamma_5
 v_{\bar p},
 \en
where $t=(p_\Lambda+p_{\bar p})^2$ and we have taken the SU(3)
symmetry result $f_3^{\Lambda p}(t)=0$ [see Eqs. (\ref{CG}) and
(\ref{vectorF})]. Since the pseudoscalar form factor $g_3$
corresponds to a kaon pole contribution to the $\Lambda\bar p$
axial matrix element, it follows that
 \be
 g_3^{\Lambda p}(t)=-{(m_\Lambda+m_p)^2\over t-m_K^2}g_1^{\Lambda
 p}(t).
 \en
Consequently,
 \be
  \la \Lambda\bar p|\bar s(1+\gamma_5)u|0\ra =
  {m_\Lambda-m_p\over m_s-m_u}f_1^{\Lambda p}(t)\bar u_\Lambda
 v_{\bar p}- {m_\Lambda+m_p\over m_s+m_u}\,{m_K^2\over t-m_K^2}
 \,g_1^{\Lambda p}(t)\bar u_\Lambda\gamma_5
 v_{\bar p}.
 \en
It is easily seen that the first term on the right hand side
satisfies the relation of vector current conservation in the SU(3)
limit, while the second term respects the PCAC relation.
Therefore, the above expression has a smooth chiral behavior in
the zero light quark mass limit $m_s,m_u\to 0$. Applying equations
of motion again yields
 \be
 \la\pi^+|\bar u(1-\gamma_5)b|\ov B^0\ra &=& {m_B^2-m_\pi^2\over
 m_b}F_0^{B\pi}(t), \non \\
 \la\rho^+|\bar u(1-\gamma_5)b|\ov B^0\ra &=& 2i\,{m_\rho\over
 m_b}A_0^{B\rho}(t)(\vp^*\cdot p_B),
 \en
where use of Eqs. (\ref{formBpi}-\ref{A3}) has been made.
Therefore, the third term in Eq. (\ref{Lambdappi}) is reduced to
 \be
 \la\pi^+|\bar u(1-\gamma_5)b|\ov B^0\ra\la \Lambda\bar p|\bar s(1+\gamma_5)u|0\ra
 &=& {m_B^2-m_\pi^2\over m_b}F_0^{B\pi}(t) \bar u_\Lambda \Big[\,{m_\Lambda-m_p\over
 m_s-m_u}f_1^{\Lambda p}(t)   \non \\ &-& {m_\Lambda+m_p\over
 m_s+m_u}\,{m_K^2\over t-m_K^2}g_1^{\Lambda
 p}(t)\gamma_5\Big]v_{\bar p}, \non \\
\la\rho^+|\bar u(1-\gamma_5)b|\ov B^0\ra\la \Lambda\bar p|\bar
s(1+\gamma_5)u|0\ra
 &=& 2i{m_\rho\over m_b}A_0^{B\rho}(t)(\vp^*\cdot p_B)\bar u_\Lambda \Big[\,{m_\Lambda-m_p\over
 m_s-m_u}f_1^{\Lambda p}(t) \non \\ &-& {m_\Lambda+m_p\over
 m_s+m_u}\,{m_K^2\over t-m_K^2}g_1^{\Lambda
 p}(t)\gamma_5\Big]v_{\bar p}.
 \en

The Clebsch-Gordon coefficients for weak $\Lambda p$ form factors
are
 \be
 d^{\Lambda p}=-{1\over \sqrt{6}}, \qquad\qquad f^{\Lambda
 p}=-{3\over\sqrt{6}}.
 \en
Hence,
 \be
 f^{\Lambda p}_1(t)=-\sqrt{3\over 2}\,F_1^p(t), \qquad\quad
 f^{\Lambda p}_2(t)=-\sqrt{3\over 2}\,F_2^p(t).
 \en
In the large $t$ regime, the dominated axial form factor is
\cite{Brodsky}
 \be \label{g1Lambdap}
 g_1^{\Lambda p}(t) \to -\sqrt{3\over 2}\,G_M^p(t).
 \en
As the $\ov B^0\to n\bar p\pi^+(\rho^+)$ decays, we consider two
distinct empirical fits for the proton magnetic form factors
denoted by CHT and BKLH. Using the same running quark masses as
before  we show the results of branching ratios in Table IV with
and without the contributions from the axial form factor
$g_1^{\Lambda p}$. When including the contribution from axial form
factors we shall assume the validity of the relation
(\ref{g1Lambdap}) for all the range of $t$. We see that the
predictions are quite sensitive to the baryonic form factors
$f_i^{\Lambda p}$ and $g_i^{\Lambda p}$. It is evident from Table
IV that the factorizable contributions to $\ov B^0\to\Lambda\bar
p\pi^+(\rho^+)$ are generally smaller than $1\times 10^{-6}$,
while the current limit is $1.3\times 10^{-5}$ \cite{PDG}. Thus
far we have neglected the nonfactorizable contributions from Fig.
2(e). The corresponding pole diagrams involve $\Sigma_b^+$ and
$\Sigma^+$ poles. Unfortunately, it is not easy to evaluate the
nonfactorizable weak matrix elements. It is conceivable that the
total decay rate will be enhanced by a factor of 2. At any rate,
we conclude that the branching ratios of $\ov B^0\to \Lambda\bar
p\pi^+$ are at most on the verge of $10^{-6}$.

\begin{table}[ht]
\caption{Same as Table III except for $\ov B^0\to\Lambda\bar
p\pi^+(\rho^+)$. The axial form factor $g_1^{\Lambda p}(t)$ is
taken to be the asymptotic form given by Eq. (\ref{g1Lambdap}).}
\begin{center}
\begin{tabular}{l| c c c c }
& \multicolumn{2}{c}{$G_M^p$ (CHT)} & \multicolumn{2}{c}{$G_M^p$
(BKLH)}
\\ \cline{2-3} \cline{4-5} \raisebox{1.5ex}[0cm][0cm]{}
 & MS & BSW & MS  & BSW   \\ \hline
 $\ov B^0\to \Lambda\bar p\pi^+$ & $2.2\times 10^{-7}$ & $3.4\times
 10^{-7}$ & $8.0\times 10^{-8}$  & $1.2\times 10^{-7}$ \\
 & $2.9\times 10^{-7}$  & $4.3\times
 10^{-7}$ & $8.5\times 10^{-8}$  & $1.3\times 10^{-7}$ \\
 $\ov B^0\to \Lambda\bar p\rho^+$ & $2.3\times 10^{-7}$  & $3.3\times
 10^{-7}$ & $6.6\times 10^{-8}$  & $9.2\times 10^{-8}$ \\
 & $4.8\times 10^{-7}$  & $6.4\times
 10^{-7}$ & $8.5\times 10^{-8}$  & $1.2\times 10^{-7}$ \\
\end{tabular}
\end{center}
\end{table}

In Sec. IV.C below we shall explain why this penguin-dominated
decay does not have a large rate. In contrast, the radiative
baryonic decay $B^-\to\Lambda\bar p\gamma$ is likely to have an
appreciable decay rate for two reasons. First, the main pole
diagram for this radiative decay comes from the strong process
$B^-\to \Lambda_b\bar p$ followed by the weak radiative transition
$\Lambda_b\to\Lambda\gamma$. Since the latter is induced by the
electromagnetic penguin mechanism $b\to s\gamma$, it has a
magnitude of order $1\times 10^{-5}$ \cite{radiative}. Second, the
coupling of the $\Lambda_b$ with the $B$ meson and the nucleon is
large. Our study indicates that $\B(B^-\to\Lambda\bar
p\gamma)\approx (1\sim 5)\times 10^{-6}$ \cite{CYradiative}.
Therefore, experimentally it would be quite interesting to measure
the radiative baryonic $B$ state $\Lambda\bar p\gamma$ and compare
with $\Lambda\bar p\pi^+(\rho^+)$.

\vskip 0.6cm \centerline{ \underline{3.~~$\ov B^0\to\Sigma^0\bar
{\lowercase{p}}\pi^+(\rho^+)$ and $\ov B^0\to\Sigma^-\bar
{\lowercase{n}}\pi^+(\rho^+)$}} \vskip 0.4cm

There are several other interesting penguin-dominated modes as
listed in (\ref{penguinmodes}), for example $\ov
B^0\to\{\Sigma^0\bar p,\Sigma^-\bar
n,\Xi^-\bar\Lambda,\Xi^-\bar\Sigma^0\}\pi^+(\rho^+)$. The
calculations are very similar to that of $\ov B^0\to\Lambda\bar
p\pi^+$. The relevant Clebsch-Gordon coefficients for weak form
factors are
 \be
&& d^{\Sigma^0p}={1\over\sqrt{2}}, \qquad\quad
 f^{\Sigma^0p}=-{1\over\sqrt{2}}, \qquad d^{\Sigma^-n}=1,
 \qquad\quad
 f^{\Sigma^-n}=-1, \non \\
&& d^{\Xi^-\Lambda}=-{1\over\sqrt{6}}, \qquad
f^{\Xi^-\Lambda}={3\over\sqrt{6}}, \qquad
d^{\Xi^-\Sigma^0}={1\over\sqrt{2}}, \qquad
f^{\Xi^-\Sigma^0}={1\over\sqrt{2}}.
 \en
Hence,
 \be
&&  f_{1,2}^{\Sigma^0p}=-{1\over\sqrt{2}}(F_{1,2}^p+2F_{1,2}^n),
 \qquad f_{1,2}^{\Sigma^-n}=-(F_{1,2}^p+2F_{1,2}^n), \non \\
&& f_{1,2}^{\Xi^-\Lambda}={3\over\sqrt{6}}(F_{1,2}^p+F_{1,2}^n),
 \qquad~~ f_{1,2}^{\Xi^-\Sigma^0}={1\over\sqrt{2}}(F_{1,2}^p-F_{1,2}^n),
 \en
The dominated axial form factors in the large $t$ regime are
 \be \label{g1Sigmap}
 g_1^{\Sigma^0p}(t)\to {1\over 3\sqrt{2}}(G_M^p+6G_M^n), \qquad
 g_1^{\Sigma^-n}(t)\to {1\over 3}(G_M^p+6G_M^n).
 \en
Obviously, $A(\ov B^0\to \Sigma^-\bar
n\pi^+(\rho^+))=\sqrt{2}A(\ov B^0\to \Sigma^0\bar
p\pi^+(\rho^+))$.

It turns out that the branching ratios of $\ov B^0\to
\Xi^-\bar\Lambda(\bar\Sigma^0)\pi^+$, being of order $5\times
10^{-8}$, are even smaller than the $\Lambda\bar p\pi^+$ final
state. Therefore, only the results for $\Sigma\bar N\pi^+(\rho^+)$
are shown in Table V. We see that (i) branching ratios of
$\Sigma^-\bar n\pi^+(\rho^+)$ lie in the ranges of $(1.0\sim
2.2)\times 10^{-6}$ and $(0.6\sim 1.6)\times 10^{-6}$,
respectively. Thus the ratio of $\rho^+/\pi^+$ is not greater than
unity, contrary to the case of $\ov B^0\to p\bar n\pi^-(\rho^-)$.
(ii) The decay rate of $\Sigma^-\bar n\pi^+(\rho^+)$ is two times
as large as that of $\Sigma^0\bar p\pi^+(\rho^+)$, but the latter
will be more easy to detect experimentally.

\begin{table}[ht]
\caption{Same as Table III except for $\ov B^0\to\Sigma\bar
N\pi^+(\rho^+)$. The axial form factor $g_1(t)$ is taken to be the
asymptotic form given by Eq. (\ref{g1Sigmap}).}
\begin{center}
\begin{tabular}{l| c c c c }
& \multicolumn{2}{c}{$G_M^p$ (CHT)} & \multicolumn{2}{c}{$G_M^p$
(BKLH)}
\\ \cline{2-3} \cline{4-5} \raisebox{1.5ex}[0cm][0cm]{}
 & MS & BSW & MS  & BSW   \\ \hline
 $\ov B^0\to \Sigma^0\bar p\pi^+$ & $1.0\times 10^{-6}$ &
 $1.6\times 10^{-6}$ & $1.4\times 10^{-6}$ & $2.0\times 10^{-6}$  \\
 & $1.1\times 10^{-6}$ & $1.8\times 10^{-6}$ & $1.2\times 10^{-6}$ & $2.2\times 10^{-6}$ \\
 \quad~~$\to \Sigma^0\bar p\rho^+$ & $6.9\times 10^{-7}$ & $6.0\times 10^{-7}$
 & $1.0\times 10^{-6}$ & $1.0\times 10^{-6}$ \\
 & $1.2\times 10^{-6}$  & $1.0\times 10^{-6}$ & $1.6\times 10^{-6}$ & $1.5\times 10^{-6}$\\
 \quad~~$\to \Sigma^-\bar n\pi^+$ & $1.9\times 10^{-6}$  &
 $3.3\times 10^{-6}$ & $2.4\times 10^{-6}$ & $4.1\times 10^{-6}$   \\
 & $2.2\times 10^{-6}$  & $3.7\times 10^{-6}$ & $2.7\times 10^{-6}$ & $4.5\times 10^{-6}$  \\
 \quad~~$\to \Sigma^-\bar n\rho^+$ & $1.4\times 10^{-6}$ & $1.2\times 10^{-6}$ &
 $2.0\times 10^{-6}$ & $2.0\times 10^{-6}$ \\
 & $2.4\times 10^{-6}$ & $2.1\times 10^{-6}$ & $3.2\times 10^{-6}$
 & $3.0\times 10^{-6}$ \\
\end{tabular}
\end{center}
\end{table}

\vskip 0.6cm \centerline{ \underline{4.~~$\ov
B^0\to\eta'\Lambda\bar p$} }\vskip 0.4cm

It has been argued in \cite{HS} that $\ov B\to\eta'\B_s\ov B$
could be the most promising charmless baryonic modes; they may be
comparable to the $\eta' K$ and a crude estimate yields
$\Gamma(\ov B^0\to\eta' \Lambda\bar p)\approx
0.3\,\Gamma(B\to\eta' K)$. Of course, the study of
$\eta'\Lambda\bar p$ is much more complicated than $\eta'K$: The
factorizable amplitudes for the former involves several 3-body
matrix elements that are difficult to evaluate. Another
complication is that what is the role played by the gluon anomaly
is still controversial and not clear even for $\eta'K$ modes, not
mentioning the three-body one, $\eta' \B_s\ov\B$. A detailed study
of $\ov B\to \eta'\Lambda\bar p$ will be presented elsewhere.

\subsection{When do we have $\Gamma(\ov B\to\B_1\ov\B_2 M)>\Gamma(\ov
B\to\B_1\ov \B_2)$ ?}

As discussed in the beginning of this section, the question of why
some of three-body baryonic $B$ decays in which baryon-antibaryon
pair production is accompanied by a meson have larger rates than
their two-body counterparts can be qualitatively understood in
terms of the Dalitz plot analysis which indicates that, for
example, the diquark $ud$ has a very large invariant mass due to
the $V-A$ nature of the $b\to ud\bar u$ process
\cite{Buchalla,Dunietz96}. If the $ud$ forms a nucleon, then it
will tend to form a highly excited baryon and will be seen as $N
n\pi(n\geq 1)$. This explains why $N\ov N$ final states have small
rates, why $p\bar\Delta$ and $\Sigma\bar\Delta$ states are leading
tree-dominated and penguin-dominated two-body baryonic $B$ decay
modes, and why the three-body mode $N\ov N \pi(\rho)$ is favored
over the two-body one. From the calculations in Sections III and
IV, we can give a more quantitative statement.

The experimental fact that the penguin-dominated decay $B^-\to
p\bar p K^-$ has a magnitude larger than the two-body counterpart
$\ov B^0\to p\bar p$ can be easily explained in the language of
the pole model. The intermediate pole states are $\Lambda_b^{(*)}$
and $\Sigma_b^{(*)}$ for the above-mentioned three-body final
state and $\Sigma_b^{(*)}$ for the two-body one. First, the
$\Sigma_b$ propagator in the pole amplitude for the latter is of
order $1/(m_b^2-m^2_\B)$, while the invariant mass of the $(pK^-)$
system can be large enough in the former decay so that the
propagator in the pole diagram is no longer subject to the same
$1/m_b^2$ suppression. Second, $\Lambda_b$ (and the anti-triplet
bottom baryon $\Xi_b$) has a much larger coupling to the $B$ meson
and the light octet baryon $\B$ than $\Sigma_b$ [see Eq.
(\ref{grel1})]. These two effects will overcome the three-body
phase space suppression to render the three-body mode dominant.
The other examples in this category are $\Gamma(\ov B^0\to p\bar
n\pi^-)>\Gamma(B^-\to n\bar p)$ as shown before and
$\Gamma(B^-\to\Lambda_c\bar p\pi^-)>\Gamma(\ov B^0\to\Lambda_c\bar
p)$ as discussed in \cite{CYBbaryon}. We have shown before that
$\Gamma(\ov B^0\to n\bar p\pi^+)>\Gamma(B^-\to n\bar p)$ even
though the pole diagram for the former does not have a $\Lambda_b$
pole. This can be comprehended from the observation that the
former is dominated by the external $W$-emission contribution
governed by the parameter $a_1$, while the latter proceeds via the
internal $W$ emission process. If the aforementioned conditions
are not satisfied, then the three-body mode will not necessarily
have larger branching ratios than the corresponding two-body ones.
For example, the penguin-dominated decays $\ov B^0\to p\bar p \bar
K^0,~n\bar n\bar K^0$ proceed through the $\Sigma_b^{(*)}$ pole
only and hence their rates are suppressed. The penguin-dominated
decays $\ov B^0\to \Lambda\bar p\pi^+(\rho^+)$ are also suppressed
relative to $p\bar p K^{(*)}$ modes due to the lack of $\Lambda_b$
poles. Indeed, we found their magnitude does not exceed $1\times
10^{-6}$.

\section{Discussions and conclusions}
We have presented a systematical study of two-body and three-body
charmless baryonic $B$ decays. We first draw some conclusions from
our analysis and then proceed to discuss the sources of
theoretical uncertainties during the course of calculation.

\begin{enumerate}

 \item The two-body baryonic $B$ decay $B\to\B_1\ov \B_2$ receives main contributions
from the internal $W$-emission diagram for tree-dominated modes
and the penguin diagram for penguin-dominated processes. We
evaluate the corresponding pole diagrams to calculate the
nonfactorizable contributions. The parity-conserving baryon matrix
elements are estimated using the MIT bag model. We found that the
bag-model estimate of baryon matrix elements are about three times
as small as the previous calculation based on the harmonic
oscillator model. The predicted branching ratios for two-body
modes are in general very small, typically less than $10^{-6}$,
except for the case with a $\Delta$ resonance in the final state.
Physically, this is because the diquark system in $b$ decay has a
very large invariant mass and hence it tends to form a highly
excited baryon state such as the $\Delta$ and will be seen as
$Nn\pi(n\geq 1)$, for example. This also explains the
non-observation of the $N\ov N$ final states. We found that the
tree-dominated decay $B^-\to p \bar\Delta^{--}$ can be of order
$10^{-6}$ due to the large coupling of the $\Delta$ with the $B$
meson and the octet baryon. This charmless two-body baryonic mode
should be readily accessible by $B$ factories BaBar and Belle.

\item Owing to large theoretical uncertainties with
parity-violating matrix elements, we focus only on the
parity-conserving contributions for two-body final states.
Nevertheless, $B^-\to n\bar p$, $\ov B\to N\bar\Delta$ and
$\Sigma\bar\Delta$ are purely parity-conserving, whereas $\ov
B^0\to\Lambda\bar\Lambda$ is purely parity-violating, provided
that the quark pair is created from the vacuum with vacuum quantum
numbers ($^3P_0$ model). These features can be tested by measuring
decay asymmetries or longitudinal polarizations.

 \item Although three-body modes in general receive factorizable
contributions, not all of them are calculable in practice due
mainly to the lack of information for three-body hadronic matrix
elements. Therefore, in many cases we still have to reply on the
pole approximation to evaluate the factorizable amplitudes.

 \item For three-body modes we focus on octet baryon final states.
The tree-dominated modes $\ov B^0\to n\bar p\pi^+(\rho^+)$ have a
branching ratio of order $(1\sim 4)\times 10^{-6}$ for the $\pi^+$
production and $(3\sim 5)\times 10^{-6}$ for the $\rho^+$
production. Moreover, $\B(\ov B^0\to p\bar n\pi^-)\sim 3\times
10^{-6}$ and $\B(\ov B^0\to p\bar n\rho^-)\sim 8\times 10^{-6}$
are predicted. There are some theoretical uncertainties for the
prediction of $B^-\to p\bar p\pi^-$ and it is conjectured to have
a branching ratio of order $10^{-6}$.

 \item Assuming a monopole $q^2$ dependence
for heavy-to-light baryon form factors, we predict that $\B(B^-\to
p\bar p K^-)\sim 4\times 10^{-6}$ and the other penguin-dominated
decays $B^-\to p\bar p K^{*-}$, $\ov B^0\to p\bar n K^-$ and $\ov
B^0\to p\bar n K^{*-}$ all have the branching ratio of order
$2\times 10^{-6}$ and their $N\bar N$ mass spectra peak at low
mass. The first one is consistent with the recent measurement of
$B^-\to p\bar p K^-$ by Belle. Therefore, several $B\to N\ov N
K^{(*)}$ decays should be easily seen by $B$ factories at the
present level of sensitivity. The study of the differential decay
rate of $B^-\to p\bar p K^-$ clearly indicates a threshold baryon
pair production and a fast recoil meson accompanied by a low mass
baryon pair.

\item The predictions of tree-dominated decays $\ov B\to p\bar
p/n\bar p,~\ov B\to N\bar\Delta$ and penguin-dominated modes $\ov
B\to \Sigma \bar p,~\Sigma\bar\Delta$ in the QCD sum-rule approach
and the diquark model are quite different from the present work.
Measurements of the above-mentioned modes can differentiate
between the different approaches.

\item The factorizable contributions to the penguin-dominated
decays containing a strange baryon, e.g., $\ov B^0\to\Sigma^0\bar
p\pi^+(\rho^+),~\Sigma^-\bar n\pi^+(\rho^+),~\Lambda\bar
p\pi^+(\rho^+)$, are calculable. While the $\Sigma\bar N\pi^+$
state has a sizable rate, of order $(1-3)\times 10^{-6}$, the
branching ratios of $\ov B^0\to\Lambda\bar p\pi^+(\rho^+)$ are in
general smaller than $10^{-6}$.

 \item Some of charmless three-body final states have a
larger rate than their two-body counterparts because (i) the
propagator in the pole diagrams for the three-body final state is
not suppressed by $1/m_b^2$, and (ii) in general the pole diagram
of the former contains a $\Lambda_b$ or $\Xi_b$ intermediate state
which has a large coupling to the $B$ meson and the light baryon,
for example $\Gamma(\ov B^0\to p\bar n\pi^-)>\Gamma(B^-\to n\bar
p)$, $\Gamma(B^-\to p\bar p K^-)>\Gamma(\ov B^0\to p\bar p)$, or
(iii) some three-body baryonic decays are dominated by the
factorizable external $W$-emission governed by the parameter
$a_1$, for example, $\Gamma(\ov B^0\to n\bar p\pi^+)>\Gamma(B^-\to
n\bar p)$.

\end{enumerate}

Needless to say, the calculation of baryonic $B$ decays is rather
complicated and very much involved and hence it suffers from
several possible theoretical uncertainties. Though most of them
have been discussed before, it is useful to make a short summary
here: (i) Since it is very difficult to evaluate nonfactorizable
and even some of factorizable amplitudes, we have relied on the
pole approximation that, at the hadron level, these amplitudes are
manifested as the pole diagrams with low-lying one-particle
intermediate states. We use the bag model to evaluate the baryon
matrix elements. Owing to the technical difficulties and the
unreliability of the model for describing negative parity
resonances, we limit ourselves to ${1\over 2}^+$ poles and hence
consider only parity-conserving amplitudes. In the future we need
a more sophisticated method to evaluate both PC and PV weak baryon
matrix elements. Another important issue is that the intermediate
pole state may be far from its mass shell and this will affect the
applicability of the quark-model estimate of baryonic matrix
elements. (ii) We have applied the $^3P_0$ quark-pair-creation
model to estimate relative strong coupling strengths. This amounts
to treating the strong $B\B_b\B$ coupling as point-like or
assuming its relative magnitude not being affected by the momentum
dependence. However, it is not clear to us how good this
approximation is.  In the future, it is important to have a solid
pQCD analysis to understand this issue. (iii) Heavy-to-light
baryon form factors are evaluated in the non-relativistic quark
model at zero recoil. However, their $q^2$ dependence is basically
unknown. We have resorted to the pole dominance approximation by
assuming a simple monopole or dipole momentum dependence. The
unknown momentum dependence for baryon form factors is one of the
major theoretical uncertainties. (iv) We have applied SU(3)
symmetry to relate the octet-octet baryonic vector form factors to
the magnetic and electric form factors of the nucleon.
Experimentally, one certainly needs measurements of nucleon
(especially neutron) electromagnetic form factors for a large
range of $q^2$. Theoretically, it is important to know how
important the SU(3) breaking effect is and how to treat the
baryonic axial form factors. (v) The three-body decays usually
proceed through several quark diagrams. To simplify the
calculation and to catch the main physics, we have often focused
only on the leading factorizable quark diagrams. It remains to
investigate nonfactorizable contributions to see their relevance.

To conclude, we have pointed out several promising charmless
two-body and three-body baryonic $B$ decay modes which have
branching ratios in the range of $10^{-5}\sim 10^{-6}$ and hence
should be measurable by $B$ factories.

\vskip 3.0cm \acknowledgments  H.Y.C. wishes to thank C.N. Yang
Institute for Theoretical Physics at SUNY Stony Brook for its
hospitality. K.C.Y. would like to thank the Theory Group at the
Institute of Physics, Academia Sinica, Taipei, Taiwan for its
hospitality. This work was supported in part by the National
Science Council of R.O.C. under Grant Nos. NSC90-2112-M-001-047
and NSC90-2112-M-033-004.

\newpage
\centerline{\bf APPENDICES}
\renewcommand{\thesection}{\Alph{section}}
\renewcommand{\theequation}{\thesection\arabic{equation}}
\setcounter{equation}{0} \setcounter{section}{0} \vskip 0.5 cm
\vskip 0.3cm

\section{Baryon wave functions}
We list the spin-flavor wave functions of baryons relevant for our
purposes:
 \be \label{spin-flavor}
\Lambda_b^{\up}&=& {1\over\sqrt{6}}[(bud-bdu)\chi_A+(12)+(13)], \non \\
\Xi_b^{0 \up}&=& {1\over\sqrt{6}}[(bus-bsu)\chi_A+(12)+(13)], \non \\
\Xi_b^{'0 \up}&=& {1\over\sqrt{6}}[(bus+bsu)\chi_s+(12)+(13)], \non \\
\Sigma_b^{+\up}&=& {1\over\sqrt{3}}[buu\chi_s+(12)+(13)], \non  \\
\Sigma_b^{0 \up}&=& {1\over\sqrt{6}}[(bud+bdu)\chi_s+(12)+(13)], \non \\
\Sigma_b^{-\up}&=& {1\over\sqrt{3}}[bdd\chi_s+(12)+(13)], \non  \\
\Sigma^{+\up}&=& {1\over\sqrt{3}}[suu\chi_s+(12)+(13)], \non \\
\Sigma^{0 \up}&=& {1\over\sqrt{6}}[(sud+sdu)\chi_s+(12)+(13)],  \\
\Sigma^{-\up}&=& {1\over\sqrt{3}}[sdd\chi_s+(12)+(13)], \non  \\
\Lambda^{\up}&=& {1\over\sqrt{6}}[(sud-sdu)\chi_A+(12)+(13)], \non \\
p^{\up}&=& {1\over\sqrt{3}}[duu\chi_s+(12)+(13)], \non \\
n^{\up}&=& {1\over\sqrt{3}}[udd\chi_s+(12)+(13)], \non \\
\Delta^{++\up} &=& uuu\chi, \non \\
\Delta^{+\up} &=& {1\over \sqrt{3}}[duu\chi+(12)+(13)], \non \\
\Delta^{0\up} &=& {1\over\sqrt{3}}[udd\chi+(12)+(13)], \non
 \en
where $abc\chi_s=(2a^\dw b^\up c^\up-a^\up b^\up c^\dw-a^\up b^\dw
c^\up)/ \sqrt{6}$, $abc\chi_A=(a^\up b^\up c^\dw-a^\up b^\dw
c^\up)/\sqrt{2}$, $abc\chi=(a^\dw b^\up c^\up+a^\up b^\up
c^\dw+a^\up b^\dw c^\up)/\sqrt{3}$, and $(ij)$ means permutation
for the quark in place $i$ with the quark in place $j$. The
spin-flavor wave function of the $\Delta$ is expressed for
$S_z={1\over 2}$. The relative sign of baryon-pseudoscalar
couplings is then fixed.

\section{Baryon matrix elements in the bag model}
Some of the details for evaluating the baryon matrix elements in
the MIT bag model are already shown in \cite{CT92,CYBbaryon}. Here
we add the result for the matrix element of $(V-A)(V+A)$ current.
Consider the four-quark operator
$O=(\bar{q_a}q_b)_\vma(\bar{q_c}q_d)_\vma$. It can be written as
$O(x)=6(\bar {q_a}q_b)^1_\vma(\bar{q_c}q_d)^2_\vma$ where the
superscript $i$ on the r.h.s. of $O$ indicates that the quark
operator acts only on the $i$th quark in the baryon wave function.
Applying the relations
  \be \label{m.e.}
\la{q'}|V_0|{q}\ra &=&  u'u+v'v, \non \\    \la{q'}|A_0|{q}\ra
&=&-i (u'v-v'u)\vec{\sigma}\cdot\hat{r}, \non  \\
\la{q'}|\vec{V}|{q}\ra &=&
-(u'v +v'u)\vec{\sigma}\times\hat{r}-i(u'v-v'u)\hat{r},    \\
\la{q'}|\vec{A}| {q}\ra
&=&\,(u'u-v'v)\vec{\sigma}+2v'v\,\hat{r}\vec{\sigma}\cdot\hat{r},
\non
 \en
leads to the PC matrix elements
 \be \label{mevma}
 \int r^2dr\la q^{a}_1q^c_2|(\bar q_aq_b)_\vma^1(\bar
 q_cq_d)_\vma^2|q^b_1q^d_2\ra_{\rm PC} &=& X_1+X_2 +(X_1-X_2)
 \vec{\sigma}_1\cdot\vec{\sigma}_2-2X_1(\vec{\sigma}_1\cdot\hat
 r)(\vec{\sigma}_2\cdot\hat r)  \non \\ &=& X_1+X_2 +{1\over
 3}(X_1-3X_2)\vec{\sigma}_1\cdot\vec{\sigma}_2,
 \en
for $(V-A)(V-A)$ current, where we have used the relation
  \be
  \int d\Omega\,\hat{r}_i\hat{r}_j={\delta_{ij}\over 3}\int
  d\Omega,
  \en
and $X_1,~X_2$ are the bag integrals
 \be
 X_1 &=& \int^R_0
 r^2dr[u_a(r)v_b(r)-v_a(r)u_b(r)][u_c(r)v_d(r)-v_c(r)u_d(r)], \non
 \\
 X_2 &=& \int^R_0
 r^2dr[u_a(r)u_b(r)+v_a(r)v_b(r)][u_c(r)u_d(r)+v_c(r)v_d(r)],
 \en
with $u_q(r)$ and $v_q(r)$ being the large and small components,
respectively, of the $1S_{1/2}$ quark wave function (see
\cite{CT92,CYBbaryon} for detail). Likewise, for $(V-A)(V+A)$
current we obtain
 \be \label{mevpa}
 \int r^2dr\la q^a_1q^c_2|(\bar q_aq_b)_\vma^1(\bar
 q_cq_d)_\vpa^2|q^b_1q^d_2\ra_{\rm PC} &=& (X_1+X_2)+{1\over
 3}[X_1+X_2-2(X_1'-X_2')]\vec{\sigma}_1\cdot\vec{\sigma}_2,
 \en
with
 \be
 X_1' &=& \int^R_0
 r^2dr[u_a(r)v_b(r)+v_a(r)u_b(r)][u_c(r)v_d(r)+v_c(r)u_d(r)], \non
 \\
 X_2' &=& \int^R_0
 r^2dr[u_a(r)u_b(r)-v_a(r)v_b(r)][u_c(r)u_d(r)-v_c(r)v_d(r)].
 \en

For numerical estimates of the bag integrals, we shall use the bag
parameters
 \be
&& m_u=m_d=0,~~~m_s=0.279\,{\rm GeV},~~~m_c=1.551\,{\rm
GeV},~~~m_b=5.0\,{\rm GeV},  \non \\    &&
x_u=2.043\,,~~~x_s=2.488\,,~~~x_c=2.948\,,~~~x_b=3.079\,,~~~
R=5.0\,{\rm GeV}^{-1}.
 \en


\end{document}